\documentclass[superscriptaddress,floatfix,aps,pre,twocolumn,notitlepage,shortbibliography,nofootinbib]{revtex4-2}

\usepackage[utf8]{inputenc}
\usepackage{amsmath}
\usepackage{amssymb}
\usepackage{refcount}
\usepackage{siunitx}
\usepackage{graphicx}
\usepackage{hyperref}
\hypersetup{colorlinks,allcolors=black}

\usepackage{natbib}
 \usepackage[capitalise]{cleveref}
\usepackage{microtype}
\usepackage{bm}
\usepackage{lipsum}
\usepackage[english]{babel}
\usepackage{color}
\usepackage{graphicx}
\usepackage{xcolor}
\usepackage{braket}

\newcommand{\red}[1]{\begin{color}{black} #1 \end{color}}

\newcommand{\abs}[1]{\left| #1 \right|}

\begin{document}
\author{Haidar Al-Naseri}
\email{haidar.al-naseri@umu.se}
\affiliation{Stanford PULSE Institute, SLAC National Accelerator Laboratory, Menlo Park, California 94025, USA}


\author{Gert Brodin}
\email{gert.brodin@umu.se}
\affiliation{Department of Physics, Ume{\aa} University, SE--901 87 Ume{\aa}, Sweden}
\title{Probing the transition from classical to quantum radiation reaction in relativistic plasma}

\begin{abstract}
We study the transition from classical radiation reaction, described by the Landau-Lifshitz model, to the quantum mechanical regime. The plasma is subject to a circularly polarized field where the self-consistent plasma current is the source of the electromagnetic field through Ampere's law. The radiation reaction implies wave energy loss, frequency up-conversion, and a modified distribution function. Increasing the value of the quantum $\chi$-parameter, the quantum results gradually differ from the classical ones. Moreover, the deviation between models also depends on the plasma parameters, including density and temperature. We discuss the implications of our findings. 
\end{abstract}
 
\maketitle

\section{Introduction}
The interest in studying quantum electrodynamics (QED) and collective plasma physics has increased over the past few decades \cite{QED-review1,QED-review3,Di-Piazza,Fisch2024}. This growing interest has been driven by steady advancements in laser facilities \cite{li2023further}, which are enabling access to high-intensity field plasma regimes. These developments have led to significant progress in the field, including the observation of pair creation at SLAC's E-144 experiment \cite{SLAC1997} and quantum radiation reaction effects observed at Rutherford \cite{Rutherford}.
The interplay between QED effects and collective plasma dynamics is also highly relevant to extreme astrophysical environments, which may soon be explored experimentally in laboratory settings \cite{Fiuza2023}.

Charged particles accelerated by external fields, such as those from accelerators or lasers, emit photons. Depending on the energy of the emitted photons, this emission can significantly affect the dynamics of the emitting particles. Their trajectories are modified by a recoil force known as radiation reaction. The impact of radiation reaction on charged particle motion has been studied extensively over the years \cite{Landau_Lif,Jackson}.
In the classical regime, the Abraham–Lorentz–Dirac (ALD) equation is commonly employed to describe radiation reaction. However, it is known to suffer from unphysical runaway solutions. In scenarios where radiation reaction acts as a small perturbation, this issue can be avoided, leading to the well-known Landau–Lifshitz (LL) equation; for further details, see \cite{Burton}. The LL equation has been solved exactly in certain restricted field configurations for the single-particle case \cite{Bulanov}.
In plasma physics, the LL equation has been implemented in Particle-In-Cell (PIC) codes, both in its classical form \cite{Silva} and in quantum-corrected generalizations \cite{Wallin,Silva_Q}. While the LL equation is effective in modeling many scenarios involving laser–plasma interactions, it is only valid in the regime where the quantum parameter $\chi$ is small, where
$\chi$  is
\begin{equation*}
\label{CHIGeneral}
    \chi=\frac{\gamma}{E_{cr} } \sqrt{|\mathbf{E}+\frac{\mathbf{v}}{c}\times \mathbf{B}|^2-\Big(\frac{\mathbf{v}}{c}\cdot \mathbf{E}\Big)^2}
\end{equation*}
where \red{$E_{cr}=m^2c^3/\abs{e}\hbar$} is the Schwinger critical field, $m$ and $e$ are the electron mass and charge, respectively,  $\hbar$ is the reduced Planck constant, $c$ is the speed of light in vacuum, and, finally, $\gamma$ is the Lorentz factor.
For the regime of $\chi \sim 1$, one should use a quantum model of radiation reaction \cite{Di-Piazza,Fisch2024}. This regime is accessible in current laser-facilities, such as the E320 at SLAC \cite{E320} and the LUXE project at DESY \cite{LUXE}.

It is not immediately clear at what values of the quantum parameter $\chi$ the Landau–Lifshitz (LL) model becomes invalid, as the formal condition $\chi \ll 1$ is rather imprecise. In this work, we investigate the transition from classical to quantum radiation reaction in the interaction between circularly polarized electric fields and a relativistic plasma. The relativistic Vlasov equation is employed to describe the interaction between the plasma dynamics and the circular fields to leading order.
To next order, the recoil force on the plasma dynamics due to radiation reaction is included. The latter mechanism is described using two separate models, firstly, using the classical Landau-Lifshitz expression (see e.g. \cite{Silva}), and, secondly, using
the quantum model introduced by Ritus \cite{Ritus1985}.  Both models lead to qualitatively similar behavior when $\chi$ is small. However, as $\chi$ increases, the predictions of the classical model begin to separate from those of the quantum model, due to the growing significance of the stochastic nature of quantum radiation emission. Both quantum and classical radiation reaction lead to damping, wave frequency up-conversion, and modifications of the background distribution. However, when it comes to the latter process, the differences between the classical and quantum descriptions are not just quantitative, but also qualitative, for sufficient magnitude of the quantum parameter.    

The organization of this paper is as follows: In \cref{Section2}, we present the kinetic theory for a circularly polarized field, where the frequency is determined using Ampere's law. We also outline the perturbative approach used to solve both the classical and quantum radiation reaction models. In \cref{Section3}, we present the results from the numerical solution of the kinetic equation, including the effects of radiation reaction. Finally, in \cref{Summary and conclusion}, we summarize the findings and discuss the conclusions of this work.

\section{Basic equations}
\label{Section2}
For sufficiently strong electromagnetic fields, the relativistic Vlasov equation for electrons must be modified to account for quantum effects, see e.g., Refs. \cite{QED-review1,QED-review2,QED-review3, E-schwinger}. For field strengths well below the Schwinger critical field, electron-positron pair production due to the Schwinger mechanism can be neglected. However, photon emission by single electrons due to nonlinear Compton scattering may become significant in case the product $\chi A_0^2$ is not too small. Here we have introduced the "laser strength" $A_0=eE/m\omega$. Here $\omega$ is the wave frequency.
Nonlinear Compton scattering has been extensively studied in the low-density plasma regime, where the electron number density is sufficiently small that the driving electromagnetic fields can be treated as solutions to Maxwell’s equations in vacuum. In this case, the focus is primarily on the properties of the resulting emission spectra (see the recent reviews \cite{QED-review1,QED-review2,QED-review3} for comprehensive references).
However, when the electron number density is higher, plasma dynamics in strong fields becomes significantly more complex, as the self-consistent plasma currents must be accounted for.  Furthermore, if the energy radiated by the electrons is not negligible, the electron equation of motion—ordinarily governed by the Lorentz force in the external field—must be corrected to include the radiation reaction force \cite{Wallin,Silva,Silva_Q}.

\subsection{Kinetic theory for circular polarized fields}
We describe the plasma dynamics using the relativistic Vlasov equation, which is modified by an additional term accounting for radiation reaction. 
Radiation reaction corrections to the Vlasov equation were first derived by Hakim et al.\ in the context of kinetic theory for many-particle systems \cite{Hakim}. Subsequent derivations of radiation reaction–corrected Vlasov equations can be found in Refs.~\cite{Burton,Kunze,Elskens}. These works primarily focused on formulating the kinetic evolution equations, rather than exploring the resulting plasma dynamics in detail. However, a kinetic study of Landau damping influenced by radiation reaction effects was presented in Ref.~\cite{Burton-2}. In addition, hydrodynamic models of relativistic plasmas incorporating radiation reaction have been developed; see, for example, Refs.~\cite{Mahajan1,Mahajan2,Mahajan3,Dalakishvili}.

In a general form, the correction to the Vlasov equation can be expressed as a collisional contribution to the evolution equation of the distribution function $f$.
\begin{equation}
\label{CircularPol}
   \bigg[ \frac{\partial }{\partial t}+\frac{\mathbf{p}}{\epsilon}\cdot \nabla_r\bigg]f
    -  e\bigg( \mathbf{E} + \frac{c\mathbf{p}}{\epsilon}\times \mathbf{B} \bigg)\cdot \nabla_p f=C_e(f)
\end{equation}
where $\epsilon=\sqrt{m^2c^4+c^2p^2}$. The term \( C_e(f) \) models the effect of photon emission on the electron distribution function. This effect can be either classical, under appropriate conditions, or quantum mechanical in nature. In both cases, we will here focus on the regime where radiation reaction can be treated perturbatively; that is, where the Vlasov equation remains valid to leading order.  From now on, we will focus on a field geometry where (in an appropriate reference frame) the electromagnetic field is only electric  in nature. In this case we can ignore Faraday's law and close the system using Ampere’s law with a vanishing magnetic field, i.e. we use
\begin{equation}
\label{Ampers1}
    \partial_t{\bf E}=-4\pi c^2 e \int d^3p \frac{{\bf p}}{\epsilon}f
\end{equation}

To prepare for numerical calculation,  we use the following normalized variables
\begin{align}
\label{Normalization}
    t_n&=\omega_c t, 
    p_n= \frac{p}{mc},  
    \epsilon_n=\frac{\epsilon}{mc^2}\\
    f_n&= \frac{m^3c^3}{n_0}f,  
    E_n= \frac{eE}{mc\omega_c},
    n_{0n}=\frac{c^3}{\omega_c^3}n_0\notag 
\end{align}
where $\omega_c=mc^2/\hbar$ is the Compton frequency, $\hbar$ is (the reduced) Planck's constant, and $n_0$ is the unperturbed electron number density ($n_{0n}$ is the normalized electron number density). Note that we will drop the subscript $n$ denoting the normalized quantities in what follows to simplify the notation.

More specifically, we study a rotating electric field that satisfies the following conditions 

\begin{eqnarray}
\mathbf{E} &\mathbf{=}&E_{0 }\left[ \mathbf{\hat{x}}\cos (\omega t)+%
\mathbf{\hat{y}}\sin (\omega t)\right] \label{Edefine} \\
\mathbf{A} &\mathbf{=}&A_{0 }\left[ \mathbf{\hat{x}}\sin (\omega t)-%
\mathbf{\hat{y}}\cos (\omega t)\right]  \label{Adefine}
\end{eqnarray}
where $E_0$ is the field amplitude and $A_0=E_0/\omega$.
We will introduce canonical momentum according to

\begin{eqnarray}
    {\bf p}={\bf q}-{\bf A}
\end{eqnarray}
Hereafter, 
${\bf q}$ denotes canonical momentum, and all phase-space functions are expressed in terms of ${\bf q}$.
This will simplify the numerical solution of the Vlasov equation.
Next we take the background distribution function to be a  Maxwell–Jüttner distribution. The solution to the Vlasov equation, neglecting radiation reaction in the first order approximation, will then be of
the form $f=C\exp (-\sqrt{1+q_{x}^{2}+q_{y}^{2}+q_{z}^{2}})/E_{th})$, where $q_{x}=p_x - A_x$%
, etc., is the canonical momentum, $C$ is a constant determining the number density such that $\int d^3p f=n_0$, 
and $E_{th}=\sqrt{1+p_{th}^2}-1$ is the thermal energy, where $p_{th}$ is the thermal momentum spread. 


While the solution of the Vlasov equation in terms of canonical momentum is an exact solution, combination with Ampere's law shows that the given ansatz without harmonics of the fundamental frequency is an approximation, although for most parameters of interest, a very good one. The details of this calculation are shown in the Appendix, and here we proceed directly to the result for the nonlinear (i.e. amplitude dependent) frequency: 

\begin{eqnarray}
&&\omega ^{2}=c_{A}\int f(q)d^{3}q\left[ \frac{1}{\sqrt{1+q_{\bot
}^{2}+A_0^{2}+q_{z}^{2}-2q_{\bot }A_0\cos \varphi _{q}}}\right. 
\nonumber \\
&&\left. -\frac{q_{\bot }^{2}}{(1+q_{\bot }^{2}+A_{0}^{2}+q_{z}^{2})^{3/2}}-\frac{15}{128}\frac{q_{\bot }^{4}A_0^{2}}{%
\left[ 1+q_{\bot }^{2}+A_0^{2}+q_{z}^{2}\right] ^{5/2}}\right] 
\label{Appendix-C}
\end{eqnarray}
where $q_{\bot }^{2}=q_x^2+q_y^2$, $\cos \varphi _{q}$ is the azimuthal angle in cylindrical coordinates of canonical momentum space, and $c_A=4\pi\alpha n_{0n}$, where $\alpha$ is the fine-structure constant.

A useful feature is
that the right-hand side does not involve the frequency anymore, as long as
the wave amplitude is written in terms of $A_0^{2}$ (rather than in
terms of the electric field amplitude). Thus, if we specify an initial value of $A_{0
}$, and use the above to calculate the frequency, the initial electric field
amplitude will be the result of the frequency calculation combined with the given value of $%
\omega A_{0 }=E_{0}$. 
In normalized variables, picking $A_{0}$ is essentially the
same thing as picking the initial gamma factors for the case of interest,
with $A_{0 }$ larger than unity.

Applying the normalization on \cref{CircularPol,Ampers1} and using the circular polarization profile, we get
\begin{equation}
\label{Circ_Vlasov}
        \frac{\partial f}{\partial t_n}
    +  E_{x}\frac{\partial f}{\partial p_{x}}
      +  E_{y}\frac{\partial f}{\partial p_{y}}
    =C_{e}
\end{equation}
\begin{equation}
\label{Ampers}
    \frac{\partial E_{i}}{\partial t}=- c_A\int d^3p \frac{p_{i}}{\epsilon}f
\end{equation}
where $i=x,y$.
The radiation reaction force is considered to be weaker than the Lorentz force $F_{RR}\ll F_{L}$. By that, the collision term is a small correction, and we can use a perturbative approach, i.e., we let $f=f_0+\delta f$, where $f_0$ is a solution to the unperturbed Vlasov equation. We only follow the evolution as long as $\delta f \ll f_0$. The solution $\delta f$
 is given by
 \begin{equation}
     \frac{\partial \delta f}{\partial t}= C_e(f_0)\longrightarrow \delta f(t)= \int ^t_0 dt' C_e(f_0(t')) \label{pert-f}
 \end{equation}

\subsection{Landau-Lifshitz radiation reaction}
For the  regime of $\chi \ll 1$, assuming a sufficiently strong inequality, the radiation force is known to behave classically. Assuming the validity condition to hold, we can use the Landau-Lifshitz force \cite{Landau_Lif,Jackson,Burton}, adapted to the case of circular fields, in our case avoiding the spatial dependence and magnetic field, which results in the simplified expression 
\begin{multline}
    \mathbf{F}_{rad} =  \frac{2 \alpha}{3} 
     \Bigg[
     \Big(\epsilon \frac{\partial}{\partial t} + \frac{\mathbf{p}}{\epsilon}\cdot \mathbf{E} \Big)\mathbf{E}\\
     -\epsilon \bigg(
     E^2 
     -\Big( \frac{\mathbf{p}}{\epsilon} \cdot \mathbf{E}\Big)^2
     \bigg)\mathbf{p}
     \Bigg] \label{Kinetic-Landau}
\end{multline}
see e.g. \cite{RR2023}.
This force term is treated as a correction term to the Lorentz force $\mathbf{F}_L \rightarrow \mathbf{F}_L+\mathbf{F}_{rad}$.

The corresponding perturbation of the Vlasov distribution due to the radiation reaction $\delta f(t)$ is calculated to be
\begin{equation}
    \delta f=\int^t_0 dt' \nabla_p(\mathbf{F}_{rad} f_0)=\frac{2}{3}\alpha \int^t_0 dt'\Big(
    \beta+\rho+\sigma +\tau\Big)f_0 
    \label{deltaf}
\end{equation}
where the terms are
\begin{align*}
    \beta&=- \frac{\partial E_x}{\partial t} \bigg(\frac{q_x-A_x}{\epsilon}f_0 +\epsilon\frac{ \partial f_0}{\partial p_x} \bigg)-\frac{\partial E_y}{\partial t} \bigg(\frac{q_y-A_y}{\epsilon}f_0 +\epsilon\frac{ \partial f_0}{\partial p_y} \bigg)\\
    \rho&=E^2\bigg(3\epsilon f_0+ \frac{p^2-1}{\epsilon}f_0 +\epsilon \mathbf{p}\cdot \frac{\partial f_0}{\partial \mathbf{p}} \bigg)\\
    \sigma&=-\frac{\mathbf{p}\cdot \mathbf{E}}{\epsilon}\bigg(\mathbf{E}\cdot \frac{\partial f_0}{\partial \mathbf{p}} \bigg)\\
    \tau&=-\frac{(\mathbf{p}\cdot \mathbf{E})^2}{\epsilon} \bigg(4 f_0
    +\mathbf{p}\cdot \frac{\partial f_0}{\partial \mathbf{p}}\bigg)
\end{align*}

Next, we are interested in the total energy 
\begin{equation}
W_{totc}=\frac{1}{2}E^2+4\pi\alpha\int(\epsilon -1)f d^3p \label{W-def}
\end{equation}
including the ordered wave energy as well as the thermal energy, but subtracting the rest mass energy. Based on \cref{deltaf}, we can calculate the loss rate associated with the radiation reaction force, and the result is  
\begin{multline}
    \frac{d W_{tot c}}{dt}= \frac{8 \pi \alpha^2  }{3} \int d^3p     \Bigg[
   \mathbf{p}\cdot   \frac{\partial \mathbf{E} }{\partial t} + \Big(\frac{\mathbf{p}}{\epsilon}\cdot \mathbf{E}\Big)^2 \\
     -p^2 \bigg(
     E^2 
     -\Big( \frac{\mathbf{p}}{\epsilon} \cdot  \mathbf{E}\Big)^2\bigg)
     \Bigg]f_0 \label{class-loss}
\end{multline}
The energy loss rate based on this expression will be compared with the corresponding loss rate in the quantum case, see the next sub-section. 

\subsection{Quantum radiation reaction}
In the regime where $\chi\ll 1$ no longer applies, the stochastic nature of photon emission becomes prominent. Accordingly, it is no longer possible to use the LL-model model for the radiation reaction force. We use instead the probability function of emitting one photon, see e.g. Refs. \cite{Ritus1985,Di-Piazza}
\begin{equation}
\label{Ritus}
        \frac{dP}{du}= -\frac{e^2 m^2c }{4 p^2 (1+u)^2}
    \Big[
    \int_{z}^{\infty} d\rho A_i(\rho)
    +\psi A_i'(z)
    \Big]
\end{equation}
where 
\begin{equation}
    \psi = \frac{2}{z}\bigg[1+\frac{u^2}{2(1+u)}\bigg]
\end{equation}
and 
\begin{equation}
    z= \bigg(\frac{u}{ \chi}\bigg)^{2/3}, u=\frac{p'-p}{p}
\end{equation}
where $p$ is the momentum of the electron after scattering and $p'$ is before scattering. Note that $A_i$ and $A_i'$ are the Airy function and its derivative.
The quantum parameter $\chi$, defined in \cref{CHIGeneral}, in the case of circular fields is
\begin{equation}
    \chi=\sqrt{\gamma ^2 E^2- (\mathbf{q}\cdot\mathbf{E})^2}
\end{equation}
where 
\begin{equation}
    \gamma= \sqrt{1+ (q_x-A_x)^2+(q_y-A_y)^2+ q_z^2}
\end{equation}
This probability is used to model the radiation reaction as
\begin{equation}
\label{Quantu_coll}
    C_{en}= -f_0 \int ^p_0 dk\frac{p}{(p-k)^2}\frac{dP}{du}+  \int_{p}^{\infty} dp'\frac{p}{p^2} f_0(p') \frac{dP}{du}
\end{equation}
where $k= p'-p$, is the energy of the emitted photon.
and $(dP/du) dpdt$ is the probability for an electron with energy $\epsilon'$ in a time interval $dt$ emitting a photon  such that we have an electron with energy in the range $(p, p+dp)$. Note that the emission probability is determined by local field values, i.e. no temporal or spatial dependence of the external field is included \cite{QEDProbabilities}.

Considering the $\chi \ll 1$-limit in \cref{Ritus} (still valid up to $\chi =0.1$ and include more physics than the Landau-Lifshitz model, see \cite{Di-Piazza}), we expand the equation in $\chi$ and keeping up to third order in $\chi$, we get
\begin{equation}
\label{Quantum_radiation reaction}
    C_{en}= \frac{2\alpha}{3}\bigg[-\frac{d}{dp}\Big(A f\Big)+\frac{1}{2}\frac{d^2}{dp^2}\Big(Bf\Big)\bigg] 
\end{equation}
where
\begin{align}
    A&=-\chi^2\bigg( 1-\frac{55 \sqrt{3}}{16} \chi \bigg)\\
    B&=\frac{55 }{16\sqrt{3}}p\chi^3
\end{align}
The first term of $A$ is in agreement with the equation for the Landau-Lifshitz radiation reaction, \cref{Kinetic-Landau}, when omitting the first term of that equation containing a temporal derivative. The term A affects the plasma by causing it to cool over time. In contrast, under the influence of B, the plasma tends to spread in momentum space, leading to a heating effect.

 We can use \cref{Quantum_radiation reaction} to model radiation as a collision operator in the relativistic Vlasov equation. We can also consider $C_{en}$ as a perturbation where we use $f_0$ to compute $\delta f$.
  \begin{equation}
 \delta f(t)=\frac{2\alpha}{3} \int ^t_0 dt'  \bigg[-\frac{d}{dp}\Big(A f_0\Big)+\frac{1}{2}\frac{d^2}{dp^2}\Big(Bf_0\Big)\bigg] \label{Q-pert}
 \end{equation}
 Based on this expression, the energy loss rate for the quantum case, corresponding to the classical energy loss rate, \cref{class-loss}, is  

\begin{multline}
    \frac{d W_{totq}}{d t}=-\alpha^2 
    \int d^3p \frac{p^2+2\epsilon^2}{\epsilon p}
    \bigg[
    \frac{2}{3}\chi^2\bigg(1-\frac{55\sqrt{3}}{16}\chi \bigg)f_0\\
    +\frac{55}{48 \sqrt{3}}\frac{d}{dp} \Big(p\chi^3f_0\Big)
    \bigg] \label{q-loss}
\end{multline}
Note that even though, the expressions for the classical and quantum loss rates, \cref{class-loss} and \cref{q-loss}, naturally differ, the definitions of the total energy (i.e. \cref{W-def}) coincide.   


\section{Numerical results}
\label{Section3}
\subsection{Preliminaries}
The kinetic system of Vlasov corrected by the Landau-Lifshitz force and the quantum model \cref{Quantum_radiation reaction} has been solved numerically for a three dimensional momentum dependence, applying spherical momentum variables ${p,\theta,\varphi}$. Note that the spatial dependence is not considered, and hence, the wave magnetic field is zero. This can correspond to a variety of setups. The first and simplest possibility is that we have a circularly polarized plasma oscillation with an infinite wavelength in the lab-system. The second case is that we have an electromagnetic wave of constant amplitude in the lab system, but we study the system in the group velocity frame (moving with a velocity $kc^2/\omega$), such that the magnetic field and spatial variations both vanish. Thirdly, the calculation can be considered as a local approximation for the case of two counter-propagating circularly polarized electromagnetic waves, where interference makes the magnetic field vanish locally.  

We are using a perturbative approach where the correction to the initial distribution function $\delta f$ is solved perturbatively, with $\delta f \ll f$. The perturbed distribution function $\delta f$ is used to express the change in different physical quantities that results from the radiation reaction. This includes the change of the temperature $\delta T$, the vector potential $\delta A$ and the wave frequency $\delta \omega$.

Since radiation reaction is of most interest in the strongly relativistic regime, corresponding to $A\gg1$, even for relativistic temperatures $T\sim 1$ (or somewhat larger), we will focus on a regime where the wave energy density is much larger than the thermal energy density. In this case, the energy radiated is matched to leading order by a decrease in the wave energy density. In order to study the wave energy loss, let us first derive the total energy loss due to radiation emission. 

The total energy in the system, including the initial distribution function $f_0$ and the perturbed one $\delta f$ is
\begin{equation}
\label{TotalEnergy}
    W_{tot}= \frac{E^2}{2}+ 4 \pi \alpha \int d^3p \epsilon \,(f_0+\delta f) 
\end{equation}

Applying a time derivative on the total energy, we get
\begin{multline}
   \frac{dW_{tot}}{dt} = E\frac{d E}{dt} +4\pi\alpha \int d^3p\epsilon\Big(\mathbf{E}\cdot \nabla_p (f_0+\delta f)+\frac{d \delta f}{dt} \Big)\\
   =4\pi\alpha \int d^3p \epsilon C_{en}{dt} \label{Wtot-rate}
\end{multline}
where we used a cancellation of the first term and the first part of the second term. Perturbatively, to leading order, $C_{en}$ can be evaluated as $C_{en}(f_0)$, where we either use the LL-expression \cref{Kinetic-Landau} or the quantum model \cref{Quantum_radiation reaction} for the radiation reaction. 

Next, we focus on the expression for the kinetic energy, and the regime with 
$A_0\gg T$. Note that this can hold even if we allow for relativistic
temperatures, of the order of the rest mass energy or larger.  We can then
expand the kinetic energy according to 
\begin{eqnarray}
W_{k} &=&\int (\epsilon -1)fd^{3}q ´\nonumber \\
&=&\int \left( \sqrt{1+A_0^{2}+q_{\bot }^{2}-2A_0q_{\bot }\cos
(\varphi _{q}-\omega t)}-1\right) fd^{3}q  \nonumber \\
&\simeq &\int A_0\left[ 1+\frac{1}{2}\left( \frac{1+q_{\bot }^{2}}{%
A_0^{2}}\right. \right. \nonumber \\
&&\left. \left. -\frac{2A_0q_{\bot }\cos (\varphi _{q}-\omega t)}{%
A_0^{2}}\right) -1\right] fd^{3}q
\end{eqnarray}%
The energy variations $\delta W_k$ may arise either from gradual changes in the vector potential amplitude $\delta A_0$ or from slow modifications of the distribution function $\delta f$. For typical values ($A_0>10$, and a
temperature $T<1$ ),the term $\propto (1+q_{\bot }^{2})/A_0^{2}$ is of the order of one percent or smaller than the leading term. Moreover, from
particle conservation $\int \delta fd^{3}q=0,$ and thus for the last term, changes in $\delta f$ will not contribute to changes in the kinetic energy.
Similarly, particle conservation ensures that changes in the energy of the
leading term correspond to changes in $\delta A_0$ only. For the
term $\propto \cos (\varphi _{q}-\omega t)$, changes in $\delta f$ will be
associated with energy changes, provided there is an anisotropic
contribution to $f$, which compensates for the dependence on $\varphi _{q}-$ in
the rest of the integrand. As we will see, radiation reaction may create
anisotropies in an initially isotropic distribution. However, since the
distribution function evolves slowly, such energy changes will have an
oscillatory dependence, reducing the importance of the term in the long-term
evolution. Moreover, the term is already smaller than the leading one by a
factor $1/A_0.$  Overall, we can therefore conclude that changes in the kinetic energy will
predominantly be associated with changes in the vector potential

The energy changes $\delta W_{k}$ may be due to slow changes in the
vector-potential amplitude $\delta A_0$, or due to changes in the
distribution function $\delta f$.  For typical values ($A_0>10$, and a
temperature $T<1$ ),the term $\propto (1+q_{\bot }^{2})/A_0^{2}$ is of
the order of one percent or smaller than the leading term. Moreover, from
particle conservation $\int \delta fd^{3}q=0,$ \ and thus for the last term,
changes in $\delta f$ will not contribute to changes in the kinetic energy.
Similarly, particle conservation assures that changes in the energy from the
leading term correspond to changes in $\delta A_0$ only. For the
term $\propto \cos (\varphi _{q}-\omega t)$, changes in $\delta f$ will be
associated with energy changes, provided there is an anisotropic
contribution to $f$, which compensate for the $\varphi _{q}-$dependence in
the rest of the integrand. As we will see, radiation reaction may create
anisotropies in an initially isotropic distribution. However, since the
distribution function evolves slowly, such energy changes will have an
oscillatory dependence, reducing the importance of the term in the long-term
evolution. Moreover, the term is already smaller than the leading one by a
factor $1/A_0.$  As a result, changes in the kinetic energy will
predominantly be associated with changes in the vector potential.

Based on the above approximation, we aim to relate the energy loss to
changes in the wave frequency and in the vector potential. Writing $A_{\bot
}=A_{0}+\delta A$ and $\omega _{0}=\omega _{0}+\delta \omega $, where index $%
0$  represents the initial value, and $\delta $ indicates changes we have%
\begin{eqnarray}
\frac{d}{dt}W_{tot} &=&\frac{d}{dt}\left[ \frac{E^{2}}{2}+4\pi \alpha \int
d^{3}q\frac{f_{0}}{\epsilon }\right]   \nonumber \\
&=&\frac{d}{dt}\left[ \omega _{0}^{2}A_{0}\left( \delta A+A_{0}\frac{\delta
\omega }{\omega _{0}}\right) \right.   \nonumber \\
&&\left. +4\pi \alpha \delta A\int d^{3}p\frac{A_{0}}{\epsilon }f_{0}\right] 
\end{eqnarray}
From the dispersion relation, $\delta A$ and $\delta \omega $ are related by 
\begin{equation}
\delta \omega =-\frac{\delta A}{2\omega _{0}}4\pi \alpha \int d^{3}p\frac{%
A_{0}}{\epsilon ^{3}}f_{0.}  \label{DR-change}
\end{equation}%
As a result, the rate of change of the vector potential can be expressed in
terms of the energy loss rate according to%
\begin{equation}
\frac{d\delta A}{dt}=\frac{1}{\omega ^{2}A_{0}-2\pi \alpha A_{0}^{3}\int
d^{3}p\frac{f_{0}}{\epsilon ^{3}}+4\pi \alpha A_{0}\int d^{3}p\frac{f_{0}}{%
\epsilon }}\frac{dW_{tot}}{dt}  \label{A-evo}
\end{equation}%
Thus, given the energy loss rate Eq. (\ref{Wtot-rate}), we can determine the decrease in
vector potential from Eq. (\ref{A-evo}), which gives the increase in
frequency from Eq. (\ref{DR-change}). These combined changes then determine the decrease rate in the wave electric field. The concrete expression for the energy loss rate varies depending on the model. Specifically, the energy loss rate due to radiation reaction for the LL-model, the LL-limit of the quantum model, and the quantum model are given by
\begin{align}
\label{Radiation_energy}
    \frac{d W_{tot,LL}}{dt}&=\frac{8}{3}\pi\alpha^2 \int d^3p \epsilon (\beta +\rho+\sigma+\tau  )f_0\notag \\
    \frac{d W_{tot,QL}}{dt}&=\frac{8}{3}\pi\alpha^2 \int d^3p \epsilon \frac{d}{dp} (-Af_0)\\
     \frac{d W_{tot,Q}}{dt}&=\frac{8}{3}\pi\alpha^2 \int d^3p \epsilon \bigg[ \frac{d}{dp} (-Af_0)
     +\frac{1}{2} \frac{d^2}{dp^2}(Bf_0)\bigg]\notag 
\end{align}

Next, we focus on changes of the background distribution. It has been noted in the literature that both cooling (see e.g. \cite{Mahajan3}) and heating (see e.g. \cite{Di-Piazza}) of the background distribution are possible. Although these changes might be of minor importance for the overall energy balance, modifications of the background distribution still constitute a prominent feature of radiation reaction. The initial temperature of the plasma 
\begin{equation}
    T_0=4\pi\alpha \frac{2}{3}\int d^3p (\epsilon_q-1)f_0
\end{equation}
where $\epsilon_q=\sqrt{1+p^2}$. For the change of the plasma temperature due to radiation, we define this with the average motion of the plasma subtracted, that is $\delta T$ is given by 
\begin{equation}
    \delta T=4\pi\alpha \frac{2}{3}\int d^3p (\epsilon_q-1)\delta f \label{delta-T}
\end{equation}
where the perturbed distribution function can be computed from \cref{pert-f}, with the concrete expressions for the classical and quantum case given by \cref{deltaf} and \cref {Q-pert}, respectively. Besides an overall change in the thermal spread of the distribution function (which can be either cooling or heating) as given by \cref{delta-T}, we will see below that more general changes than cooling/heating in the shape of the background distribution are also possible.  

\begin{figure}
    \centering
    \includegraphics[width=\linewidth]{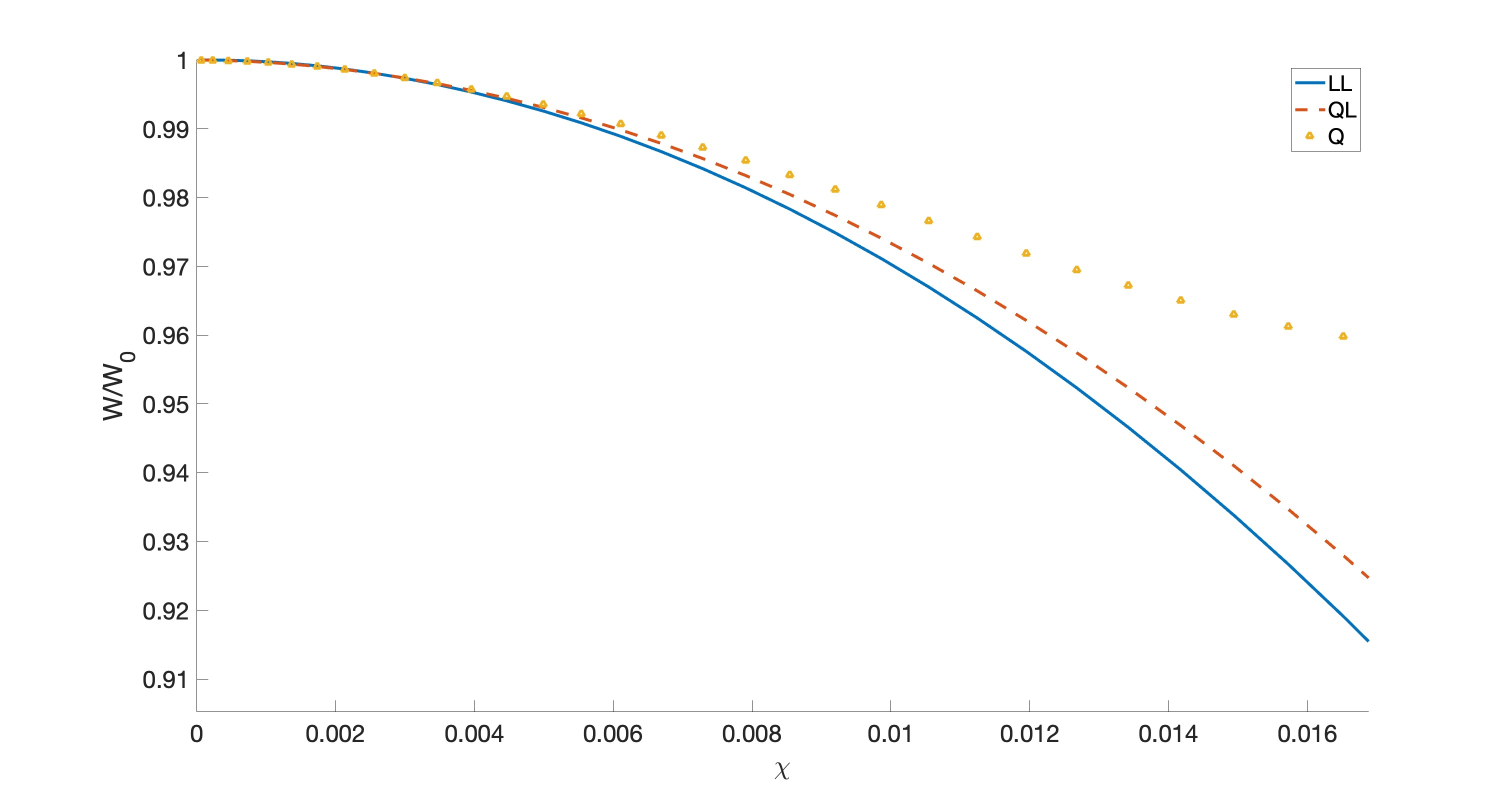}
    \caption{The total energy according to \cref{TotalEnergy}, divided by the initial total energy $W_0=W(t=0)$ is plotted versus $\chi$ using three models, quantum (Q), LL-limit of the quantum model (QL) and Landau-Lifshitz (LL) model. Here we have used a plasma density $n_0=7\times 10^{21}/cm^3$ and $p_{th}=1$.}
    \label{Fig1}
\end{figure}

\subsection{Accuracy of the Landau-Lifshitz expression}
As is well known, the Landau-Lifshitz radiation reaction model predictions are valid for $\chi \ll 1$. Theoretically, it is proven that the quantum radiation reaction model agrees to all orders with the Landau-Lifshitz model for very small $\chi$ \cite{Di-Piazza}.
However, for $\chi$ approaching unity, the predictions from the Landau-Lifshitz model are not at all accurate, and one should use the quantum model. However, a more precise assessment than this would be desirable, accurately specifying how large the values of $\chi$ can be before the Landau-Lifshitz model starts to deviate significantly from the quantum one. This is what we will be looking for in this subsection.
Studying the equation for the quantum model \cref{Quantum_radiation reaction} in detail, we can see that there are two types of effects contributing with different signs. It is shown in Ref. \cite{Di-Piazza} that by dropping the term $\propto\chi^3$ in \cref{Quantum_radiation reaction}, one gets the equation of Landau-Lifshitz. The higher order term dominates for large $\chi$ and/or large gamma-factors.

We start our analysis by looking at the radiation energy loss, using the expressions in \cref{Radiation_energy}. In \cref{Fig1} we plot the total energy of the plasma and the field divided by the initial energy $W_0$ in the system using the classical Landau-Lifshitz (LL), the Landau-Lifshitz limit of the quantum model (QL) (obtained by omitting the $\chi^3$-term in \cref{Quantum_radiation reaction}), and the quantum model (Q) as a function of $\chi$. We can clearly see that all three models have almost the same radiation energy for small $\chi$ as expected. For larger $\chi$, in particular $\chi >0.01$, the quantum model predicts a slower increase of the radiation energy compared to the other models. This has to do with the different signs of the two terms in \cref{Quantum_radiation reaction} that represents cooling and heating of the plasma. For larger $\chi$, the $\chi^3$-terms start to be important and reduces the effect of the $\chi^2$-term, that otherwise keeps increasing for larger $\chi$ in the same way as classical Landau-Lifshitz. Simulations were terminated at the maximum $\chi$-value where the perturbative solution remains valid.

\begin{figure}
    \centering
    \includegraphics[width=10 cm,height=10 cm]{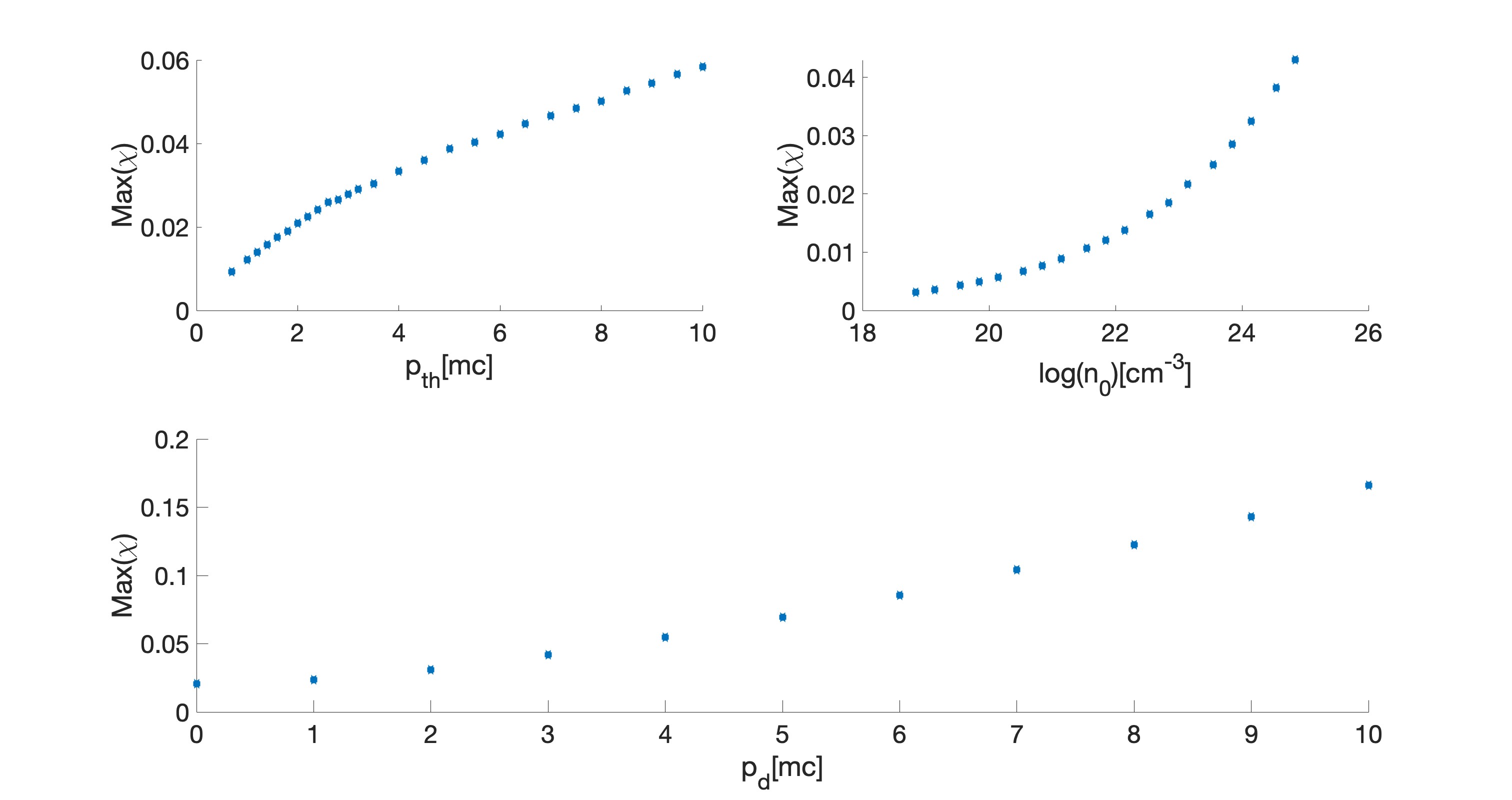}
    \caption{We demonstrate the validity of the classical LL-model by plotting the maximum value of $\chi$ for which the classical model remains applicable. In the upper-left panel, we show the validity as a function of the initial plasma thermal spread $p_{\text{th}}$. The upper-right panel presents the validity versus the plasma density $n_0$, while the lower panel illustrates the validity as a function of the radial drift in momentum space, $p_d$.}
    \label{Fig2}
\end{figure}

The maximum $\chi$-value for which predictions from the LL-model are valid to a useful approximation varies depending on the input parameters of the plasma. At what $\chi$ the contribution from the higher order terms in \cref{Quantum_radiation reaction} becomes larger than $\chi^2$ depends on the temperature and the drift on the plasma as well as the plasma density. In \cref{Fig2}, we plot the maximum $\chi$ at which the LL-model is valid to a useful approximation vs. input parameters of the plasma. Here "useful approximation" is defined to apply as long as the $\chi^2$-term is larger than the $\chi^3$-term, making the quantum model still predict more cooling than heating. 
In the upper panel to the left of \cref{Fig2}, we plot the validity of the LL-model versus the thermal momentum of the plasma. A large thermal spread of the initial plasma makes it possible for the cooling part of \cref{Quantum_radiation reaction} to be dominating for a larger $\chi$. As can be expected, the hotter the plasma, the more likely it is that the cooling dominates over the heating mechanism. Hence, the LL-model is valid for a larger $\chi$ when the temperature of the plasma is higher. 

For the right upper panel of \cref{Fig2}, we plot the validity of the LL-model vs. the plasma density.
The quantum parameter $\chi$ can effectively be defined as $\chi=\omega A_0^2$. For higher plasma densities, the frequency $\omega$ is higher, leading to a lower value of $A_0$ (average gamma factor) for a given value of $\chi$. Since the dominant part of the diffusive term receives an extra factor of the order $p/p_{th}\sim A_0/p_{th}$, a higher density (and thereby reduced $A_0$) leads to an increased validity for the classical model as a function of $\chi$.

Adding an initial radial drift $p_d$ to the initial distribution of the plasma, we replace the exponential factor in the Maxwell-Juttner distribution according to $\exp(-\sqrt{1+p^2}/E_{th}) \rightarrow \exp(-\sqrt{1+(p-p_d)^2}/E_{th})$. We investigate the validity of the LL-model as a function of drift momentum $p_d$ in the lower panel of \cref{Fig2}. The drift extends the validity of the LL-model to a larger regime of $\chi$ when the drift momentum is larger than the thermal momentum. This is because the particles in the plasma are shifted to a higher momentum, leading to a larger reservoir of energy that can be cooled. This makes the cooling mechanism dominate over the heating for a larger value of $\chi$.

\begin{figure}
    \centering
    \includegraphics[width=10 cm,height=10cm]{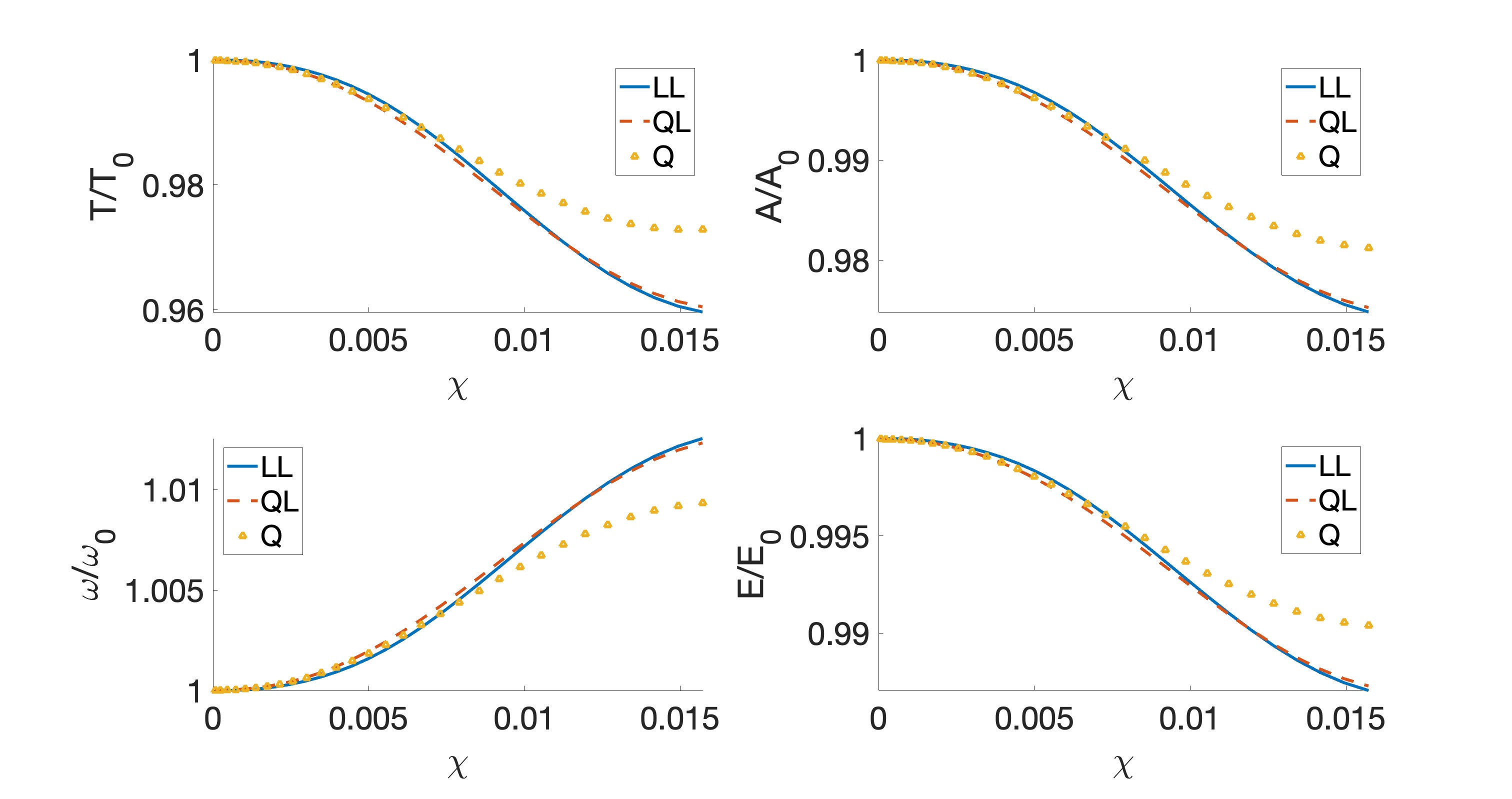}
    \caption{A comparison of the three models used in the paper. In the first subplot, we plot the temperature  $ T$ divided by the initial temperature $T_0$ vs. $\chi$. In the second subplot, we plot the cycle-average gamma-factor $ A$ divided by the initial average gamma factor $A_0$. In the third subplot, we plot the plasma frequency $ \omega$ divided by the initial plasma frequency $\omega_0$. Finally, in the last subplot, we plot the field amplitude $E$ divided by the initial amplitude $E_0$ vs. $\chi$. For all plots, we have used plasma with the density $n_0=7\times 10^{21}/cm^3$ and a thermal momentum $p_{th}=1$ (=$mc$ in dimensional units). The solid line is for the classical Landau-Lifshitz model, dashed line for the classical limit of the quantum model and dotted line for the Quantum model.  }
        \label{Fig3}
\end{figure}

\begin{figure}
    \centering
    \includegraphics[width=9 cm,height=10cm]{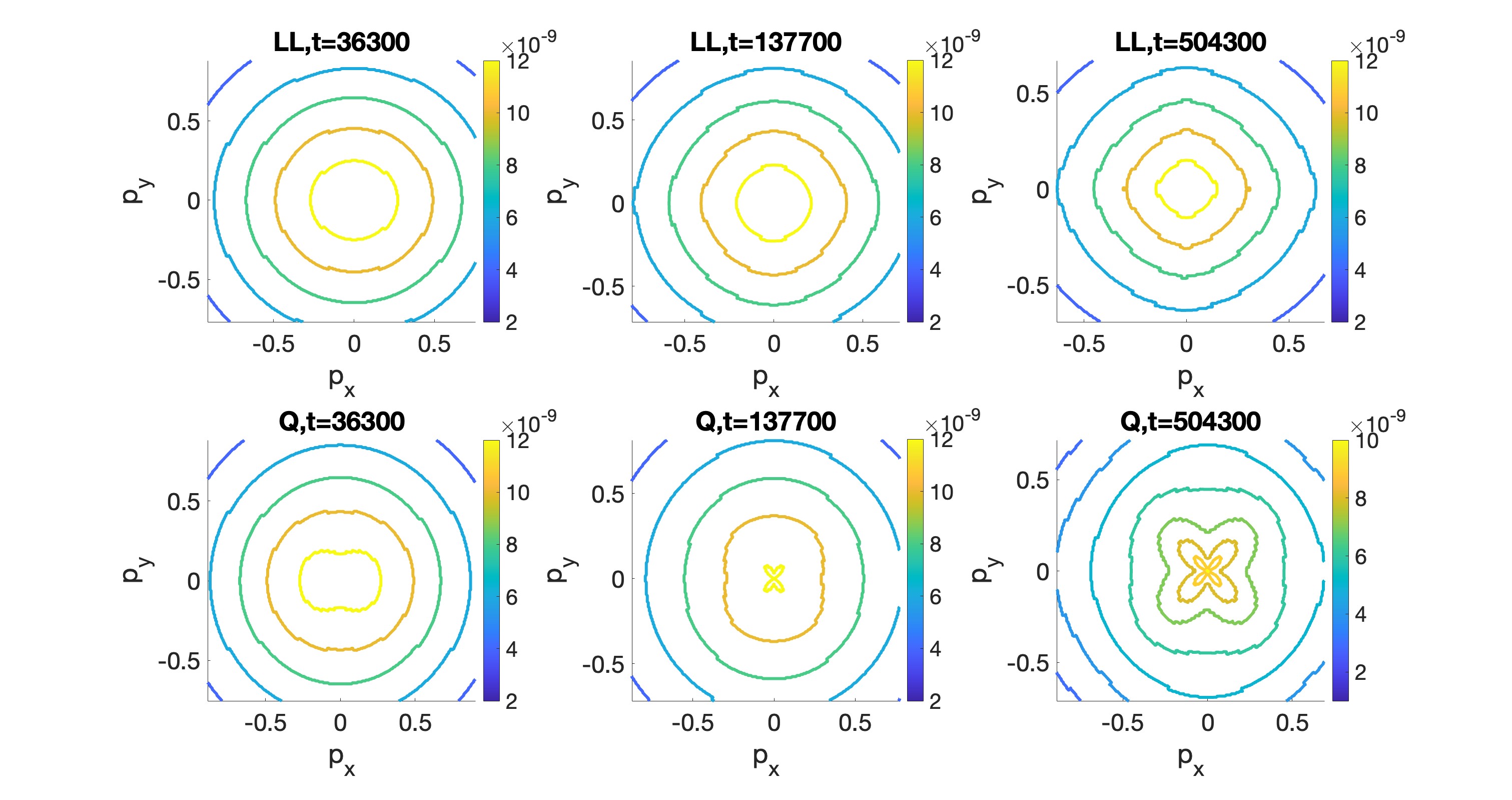}
    \caption{Contour plots of $p_x-p_y$-plane of particle distribution using both the classical LL-model and the quantum model. We have used a plasma density $n_0=7\times 10^{21}/cm^3$ and $A_0=33$.}
    \label{Fig4}
\end{figure}

Next, we study the predictions from the three models as a function of $\chi$ regarding temperature change, frequency up-conversion, etc., in \cref{Fig3}. In the panel to the left, we see how the LL-model predicts a continuous decrease in temperature. By contrast, the quantum model is beginning to show a conversion from cooling to heating for sufficiently large $\chi$. 

In the upper right panel, we plot the decrease of $A/A_0$, which is effectively the average gamma factor. Due to the radiation reaction, the average gamma factor is decreased as the kinetic energy decreases. The quantum model shows that the decrease is slowing down with increasing $\chi$, as compared to the classical result, in accordance with radiation energy loss in \cref{Fig1}.

In the lower left panel, we plot the frequency upshift due to the reduction of the gamma factor, where, in the case of larger $\chi$, the upshift is less pronounced in the quantum case. Due to the frequency upshift, the decrease in electric field is less prominent than the decrease in vector potential (average gamma factor), as shown in the lower right panel of \cref{Fig3}. Similarly to previous cases, the deviation between classical and quantum behavior comes from the diffusive term, that is, the LL-model is in good agreement with the quantum model when the diffusive term has been dropped.    

\subsection{Modification of the background distribution function}
In this subsection, we examine how radiation modifies the plasma background distribution. Since the plasma is predominantly accelerated in the \(xy\)-plane, we focus on the momentum components \(p_x\) and \(p_y\). In \cref{Fig4}, we compare the evolution of the plasma distribution under the influence of both the Landau–Lifshitz and quantum radiation reaction models. The simulation parameters are \(A_0 = 33\), a plasma density of \(n = 7 \times 10^{21}\,\mathrm{cm}^{-3}\), and a quantum parameter \(\chi = 0.0118\).

At early times (first column), both models predict similar modifications to the plasma distribution. However, already in the second column, a clear deviation emerges: the quantum model begins to exhibit qualitatively different behavior. The classical model shows a consistent shift of electrons toward lower energies, an effect that becomes especially evident when following the high-density contours (highlighted in yellow in Fig. 4) across the first and second panels of the top row. This reflects a cooling mechanism in which radiation causes a net loss of thermal energy from the plasma.
In contrast, the quantum model displays both cooling and opposing heating effects. Although some electrons still accumulate at lower momenta, the distribution also exhibits splitting behavior. This is confirmed in the third column, where the quantum simulation shows a clear bifurcation of the plasma into two distinct momentum regions. Due to this heating effect, electrons are spread over a wider momentum range, resulting in a decrease in the peak value of the distribution function, from $1.2 \times 10^{-8}$ to $10^{-8}$. This reduction arises because electrons are pushed away from the low-momentum region, lowering the amplitude of the central peak in the distribution.

\begin{figure}
    \centering
    \includegraphics[width=10 cm,height=10cm]{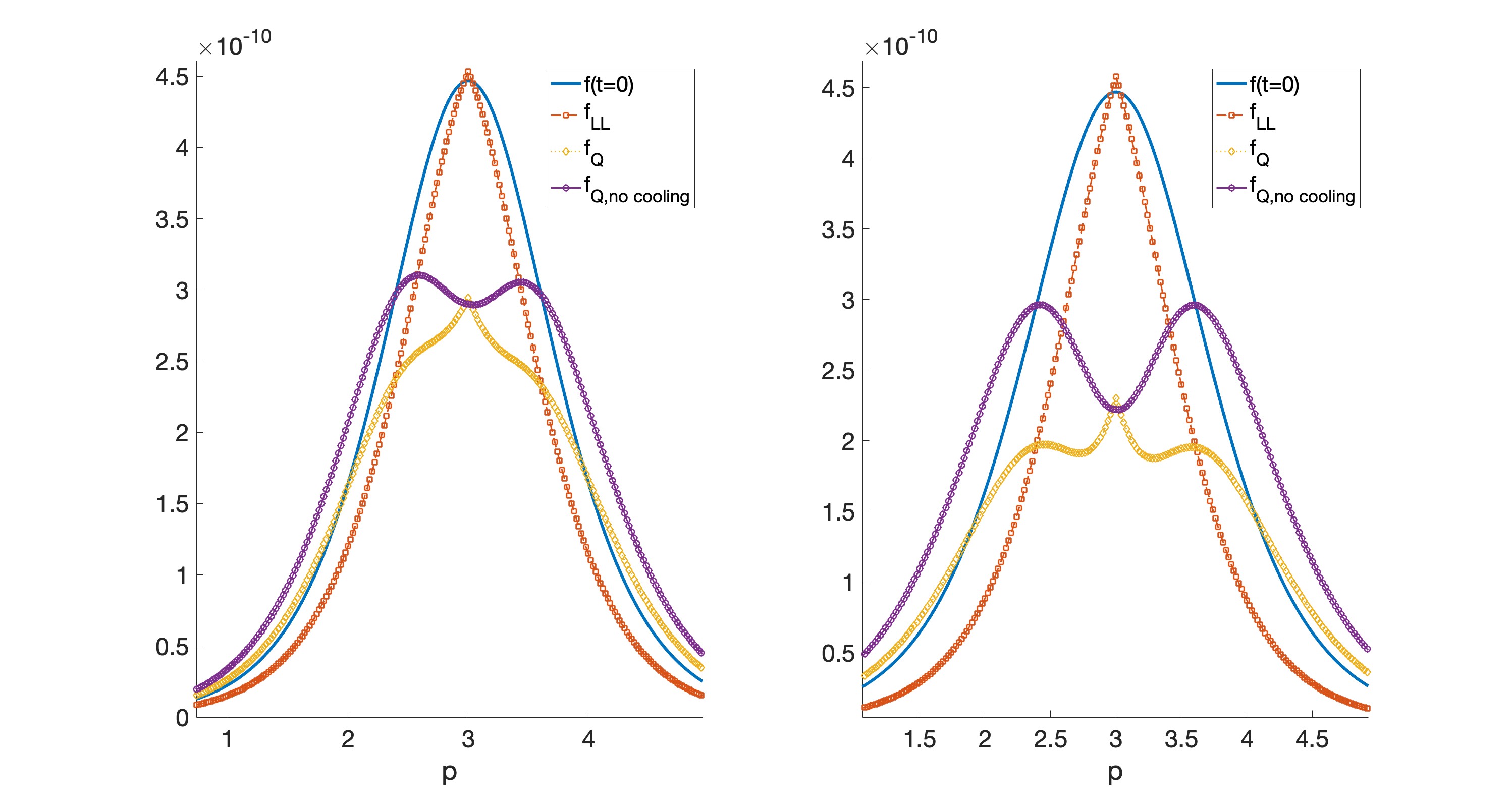}
    \caption{Radial momentum distribution of the initial plasma (solid curve), the plasma distribution for the LL-model (dashed line with squares), the plasma distribution for the model including quantum radiation (dotted line with diamonds) and the plasma distribution for the model including quantum radiation, but excluding cooling effects (solid line with circles). We have been using a plasma distribution with density $n_0=7\times 10^{21}/cm^3$, $A_0=40$, implying $\chi=0.024$, and a radial momentum drift $p_d=3$. In the left panel, we had $t=8000$ while in the right panel we had $t=20000$.   }
    \label{Fig5}
\end{figure}
To understand the mechanism behind the splitting of the plasma distribution in greater detail, we further investigate the problem in \cref{Fig5}, where we plot the radial momentum distribution of the plasma. For illustrative purposes, we initialize the plasma with a radial momentum drift $p_d = 3$.
As expected, the Landau–Lifshitz (LL) model predicts a cooling effect, where the plasma distribution is compressed toward the drift value $p_d$. As the simulation progresses, the LL distribution $f_{\mathrm{LL}}$ becomes more narrow, and the peak near $p = p_d$ becomes more pronounced.
For the quantum model, we consider two variants. The first is the full quantum distribution $f_{\mathrm{Q}}$, which includes both heating and cooling effects. The second is a modified version that excludes the cooling effect, constructed by removing the first term in \cref{Quantum_radiation reaction}, which represents the cooling mechanism.
In the left panel of \cref{Fig5}, we observe that both quantum variants give a broader plasma distribution than the initial state. The complete quantum model $f_{\mathrm{Q}}$ is somewhat narrower than the quantum model without cooling, since the inclusion of cooling partially counteracts heating, particularly near $p = p_d$. The same mechanism that causes the LL distribution to narrow and peak near $p_d$ is also responsible for reshaping the quantum distribution—without cooling—to a more focused profile around the drift momentum.
Additionally, the quantum models exhibit a tendency for the distribution to split into two distinct regions in momentum space. This effect is more pronounced in the version without cooling.
In the right panel of \cref{Fig5}, which corresponds to a later simulation time, the difference between the quantum and LL distributions becomes even more apparent. The splitting behavior in the quantum model is enhanced, further distinguishing it from the classical prediction.

\section{Summary and Conclusion}
\label{Summary and conclusion}

In the present paper we have compared classical radiation reaction with the quantum case, for the case of a rotating electric field, self-consistently sustained by circularly polarized plasma currents. The classical and quantum mechanical descriptions share a number of features, including electric field damping, frequency up-conversion, and a modification of the background distribution function (initially chosen as a Maxwell–Jüttner distribution). The electric field damping (or wave energy damping) and frequency up-conversion are qualitatively the same in the classical and quantum descriptions. For a very small $\chi$-parameter, the wave damping rate and the frequency up-conversion rate found from the different models agree. However, a gradually increasing difference can be seen roughly from $\chi\sim 0.005$ and upward. Generally, the classical model overestimates the damping rate and the frequency up-conversion, as compared to the quantum model.  A hybrid model, including the quantum modification of the friction term of the classical LL-model, but omitting the diffusive quantum term, produces results intermediate between the full quantum model and the LL-model, mostly in rather good agreement with the classical model.  In particular, when it comes to frequency up-conversion and temperature modifications, the majority of the quantum and classical disagreements come from the diffusive term, and the quantum modification of the friction term only has limited importance. 

Importantly, the level of agreement between quantum and classical models depends not only on the $\chi$-parameter but also on the plasma density and temperature.  Specifically, the value of $\chi$ needed for the classical and quantum models to show significant disagreement increases with the plasma temperature and with the plasma density.  The density dependence is due to the increase in the plasma frequency with density, which lead to a lower momentum of the particles, for the same $\chi$. 
The reason for the dependence on the thermal momentum $p_{th}$, is fairly simple, as the cooling mechanism is more prominent if the temperature is higher, making the quantum transition from cooling to heating require a higher value of $\chi$. The mechanism responsible for the density dependence of the classical-quantum transition is still present, but it will be less effective compared to temperature dependence.     

A major difference between the classical and quantum models concerns the modification of the background distribution. Previous authors \cite{Di-Piazza} have shown that the background cooling associated with the classical LL-model can be turned into heating when $\chi$ is sufficiently large to enter the quantum regime. This feature is confirmed, but we also observe the presence of anisotropic modifications of the background distribution. The anisotropic modification is a pronounced property of the quantum model, and the main contribution comes from the diffusive term of the quantum model. While this is a distinctive property of the quantum regime, it might still be difficult to confirm experimentally, as the anisotropy evolves on the rapid scale of the oscillatory field. A long-term study of the evolution of the background distribution might give improved possibilities for experimental confirmation. However, such a study is out of the scope of the present paper.     
\section{Appendix}

Here we will derive the expression for the self-consistent frequency, $%
\omega $, to be used in the unperturbed (without radiation reaction) solution to the Vlasov equation. As we will see, for a large amplitude wave
in the relativistic case, even for a circular polarized field, we will
obtain a certain degree of higher harmonic generation in the Vlasov
description. Nevertheless, for parameters of interest (in the
context of parameters relevant for radiation reaction), the harmonic
generation will be small. Thus, we can still start with a circularly
polarized ansatz without harmonic content, with an electric field and vector potential given by Eqs. (\ref{Edefine}-\ref{Adefine}). Putting these ansatzes into the Vlasov equations, using
canonical variables with $q_{x}=p_{x}+eA_{x}$, $q_{x}=p_{x}+eA_{y}$, $%
q_{z}=p_{z}$, we obtain a time-independent solution for the distribution
function $f(\mathbf{q})$. Substituting this result into Ampere's law gives 
\begin{widetext}
\begin{eqnarray}
&&-\omega ^{2}A_0\left[ \cos \left( \omega t\right) \mathbf{\hat{x}+}%
\sin \left( \omega t\right) \mathbf{\hat{y}}\right]   \nonumber \\
&=&c_A\int \frac{[q_{x}-A_0\cos \left( \omega t\right) ]\mathbf{\hat{x}+}%
[q_{y}-A_0\sin \left( \omega t\right) ]\mathbf{\hat{y}}}{\sqrt{%
1+\left\{ [q_{x}-A_0\cos \left( \omega t\right) ]\mathbf{\hat{x}+}%
[q_{y}-A_0\sin \left( \omega t\right) ]\mathbf{\hat{y}}\right\}
^{2}+q_{z}^{2}}}f(\mathbf{q})d^{3}q  \label{Ampere-A}
\end{eqnarray}%
\end{widetext}
If the thermal effects are small, such that $q_{x},q_{y}\ll A_0$, the
gamma factor in the denominator would be approximately constant, $\gamma
\simeq \sqrt{1+A_{0 }^{2}}$, and in this approximation we obtain a very
simple expression for the frequency, as the terms proportional to $q_{x}$
and $q_{y}$ vanish because they are odd functions, and the remaining
time-dependence cancels out, confirming the applicability of the ansatz.
However, in order to go beyond the lowest order approximation, and allow for higher temperatures, we expand the denominators according to 
\begin{eqnarray}
&&\frac{1}{\sqrt{1+\left\{ [q_{x}-A_0\cos \left( \omega t\right) ]%
\mathbf{\hat{x}+}[q_{y}-A_0\sin \left( \omega t\right) ]\mathbf{\hat{y}%
}\right\} ^{2}+q_{z}^{2}}}  \nonumber \\
&=&\frac{1}{\sqrt{1+q_{\bot }^{2}+A_0^{2}}\sqrt{1-\frac{(2q_{x}A_{\bot
}\sin (\omega t)+2q_{y}A_0\cos (\omega t))}{1+q_{\bot }^{2}+A_{\bot
}^{2}}}}  \nonumber \\
&\simeq &\frac{1}{\sqrt{1+q_{\bot }^{2}+A_0^{2}}}\left[ 1+\frac{1}{2}%
\frac{2q_{x}A_0\sin (\omega t)+2q_{y}A_0\cos (\omega t)}{%
1+q_{\bot }^{2}+A_0^{2}}\right.   \nonumber \\
&&\left. +\frac{3}{8}\frac{\left[ 2q_{x}A_0\sin (\omega
t)+2q_{y}A_0\cos (\omega t)\right] ^{2}}{\left[ 1+q_{\bot
}^{2}+A_0^{2}\right] ^{2}}\right.   \nonumber \\
&&\left. +\frac{3}{16}\frac{\left[ 2q_{x}A_0\sin (\omega
t)+2q_{y}A_0\cos (\omega t)\right] ^{3}}{\left[ 1+q_{\bot
}^{2}+A_0^{2}\right] ^{3}}+...\right]   \label{Appendix-B}
\end{eqnarray}%
Note that the expansion parameter is always smaller than unity, and hence executing the expansion to all orders gives a converging series.

Now let us first ignore the time-dependent terms proportional to $A_{\bot
}\sin (\omega t)$ and $A_0\cos (\omega t)$ in the numerator of (\ref%
{Ampere-A}). We note that when expanding the powers of $\left[ 2q_{x}A_{0}\sin (\omega t)+2q_{y}A_0\cos (\omega t)\right] ^{n}$, all factors
will either be proportional to odd or even powers of $q_{x}$ and $q_{y}$. Thus,
we can immediately idenify the rather limited number of terms that survives
the momentum integration. One might think that the time-dependent terms
proportional to higher powers of $\cos ^{2}(\omega t)\sin (\omega t)$, and
similar, would induce higher harmonics. However, keeping only terms up to $\ %
\left[ 2q_{x}A_0\sin (\omega t)+2q_{y}A_0\cos (\omega t)\right]
^{3}$, i.e. those written out explicitly in the expansion (\ref{Appendix-B}%
), somewhat surprisingly this will not happen. This is not a general feature
due to the symmetry of the problem, and it turns out that further terms in
the expansion eventually will induce fifth harmonics. However, up to the
given order in the expansion, no harmonics of the fundamental frequency will
be created. Specifically, this means that source terms for second and cubic
harmonic terms that seem to be present \ in the expression cancel when the
integration over momentum is carried out. Nevertheless, there are still
terms in the expansion that survive, but up to the given order, they are
consistent with the monochromatic ansatz. Still overlooking correction terms in the expansion for terms
proportional to $A_0\sin (\omega t)$ and $A_0\cos (\omega t)$ in
the numerator of (\ref{Ampere-A}), the surviving terms in the expansion described above 
modify the dispersion relation according to

\bigskip 
\begin{eqnarray}
\omega ^{2} &=&c_{A}\int \frac{f(q)d^{3}q}{\sqrt{1+q_{\bot }^{2}+A_{0
}^{2}+q_{z}^{2}}}\left[ 1-\frac{q_{\bot }^{2}}{1+q_{\bot }^{2}+A_{0
}^{2}+q_{z}^{2}}\right. \nonumber  \\ \
&&\left. -\frac{15}{128}\frac{q_{\bot }^{4}A_0^{2}}{\left[ 1+q_{\bot
}^{2}+A_0^{2}+q_{z}^{2}\right] ^{3}}\right] \label{A36}
\end{eqnarray}%
Next, we note that for the terms proportional to $A_0\sin
(\omega t)$ and $A_0\cos (\omega t)$ in the numerator of (\ref%
{Ampere-A}), we do not need to apply the expansion (\ref{Appendix-B}). The reason is that the azimuthal momentum integration remove the extra
time-dependence from the denominator. However, we note that the overall
{\it magnitude} of the harmonically oscillating terms is still affected by the
time-dependent terms in the gamma-factor. This can be seen by making a change of azimuthal momentum angle when integrating $\varphi_q\rightarrow \varphi_q+\Delta\phi$, noting that $\Delta\phi$ can swallow the effect of the time-dependence in the denominator in case we let $\Delta\phi=\omega t$. Accounting for this, the final
version of the dispersion relation becomes: 
\begin{eqnarray}
&&\omega ^{2}=c_{A}\int f(q)d^{3}q\left[ \frac{1}{\sqrt{1+q_{\bot
}^{2}+A_0^{2}+q_{z}^{2}-2q_{\bot }A_0\cos \varphi _{q}}}\right. 
\nonumber \\
&&\left. -\frac{q_{\bot }^{2}}{(1+q_{\bot }^{2}+A_{0}^{2}+q_{z}^{2})^{3/2}}-\frac{15}{128}\frac{q_{\bot }^{4}A_0^{2}}{%
\left[ 1+q_{\bot }^{2}+A_0^{2}+q_{z}^{2}\right] ^{5/2}}\right] 
\label{Appendix-C}
\end{eqnarray}%
While this dispersion relation is approximate, the terms omitted are smaller
than the dominating term by a factor of the order $q_{\bot }^{6}/\left[
1+q_{\bot }^{2}+A_0^{2}+q_{z}^{2}\right] ^{3}$. Thus, in practice, the
harmonically oscillating solution applies to a very good approximation even
if the thermal momentum spread given by $f(q)$is just slightly smaller than
the field induced momenum amplitude $A_0$.



 \bibliography{References}

\begin{thebibliography}{30}%
\makeatletter
\providecommand \@ifxundefined [1]{%
 \@ifx{#1\undefined}
}%
\providecommand \@ifnum [1]{%
 \ifnum #1\expandafter \@firstoftwo
 \else \expandafter \@secondoftwo
 \fi
}%
\providecommand \@ifx [1]{%
 \ifx #1\expandafter \@firstoftwo
 \else \expandafter \@secondoftwo
 \fi
}%
\providecommand \natexlab [1]{#1}%
\providecommand \enquote  [1]{``#1''}%
\providecommand \bibnamefont  [1]{#1}%
\providecommand \bibfnamefont [1]{#1}%
\providecommand \citenamefont [1]{#1}%
\providecommand \href@noop [0]{\@secondoftwo}%
\providecommand \href [0]{\begingroup \@sanitize@url \@href}%
\providecommand \@href[1]{\@@startlink{#1}\@@href}%
\providecommand \@@href[1]{\endgroup#1\@@endlink}%
\providecommand \@sanitize@url [0]{\catcode `\\12\catcode `\$12\catcode `\&12\catcode `\#12\catcode `\^12\catcode `\_12\catcode `\%12\relax}%
\providecommand \@@startlink[1]{}%
\providecommand \@@endlink[0]{}%
\providecommand \url  [0]{\begingroup\@sanitize@url \@url }%
\providecommand \@url [1]{\endgroup\@href {#1}{\urlprefix }}%
\providecommand \urlprefix  [0]{URL }%
\providecommand \Eprint [0]{\href }%
\providecommand \doibase [0]{https://doi.org/}%
\providecommand \selectlanguage [0]{\@gobble}%
\providecommand \bibinfo  [0]{\@secondoftwo}%
\providecommand \bibfield  [0]{\@secondoftwo}%
\providecommand \translation [1]{[#1]}%
\providecommand \BibitemOpen [0]{}%
\providecommand \bibitemStop [0]{}%
\providecommand \bibitemNoStop [0]{.\EOS\space}%
\providecommand \EOS [0]{\spacefactor3000\relax}%
\providecommand \BibitemShut  [1]{\csname bibitem#1\endcsname}%
\let\auto@bib@innerbib\@empty
\bibitem [{\citenamefont {Fedotov}\ \emph {et~al.}(2022)\citenamefont {Fedotov}, \citenamefont {Ilderton}, \citenamefont {Karbstein}, \citenamefont {King}, \citenamefont {Seipt}, \citenamefont {Taya},\ and\ \citenamefont {Torgrimsson}}]{QED-review1}%
  \BibitemOpen
  \bibfield  {author} {\bibinfo {author} {\bibfnamefont {A.}~\bibnamefont {Fedotov}}, \bibinfo {author} {\bibfnamefont {A.}~\bibnamefont {Ilderton}}, \bibinfo {author} {\bibfnamefont {F.}~\bibnamefont {Karbstein}}, \bibinfo {author} {\bibfnamefont {B.}~\bibnamefont {King}}, \bibinfo {author} {\bibfnamefont {D.}~\bibnamefont {Seipt}}, \bibinfo {author} {\bibfnamefont {H.}~\bibnamefont {Taya}},\ and\ \bibinfo {author} {\bibfnamefont {G.}~\bibnamefont {Torgrimsson}},\ }\bibfield  {title} {\bibinfo {title} {Advances in qed with intense background fields},\ }\href@noop {} {\bibfield  {journal} {\bibinfo  {journal} {arXiv preprint arXiv:2203.00019}\ } (\bibinfo {year} {2022})}\BibitemShut {NoStop}%
\bibitem [{\citenamefont {Gonoskov}\ \emph {et~al.}(2022)\citenamefont {Gonoskov}, \citenamefont {Blackburn}, \citenamefont {Marklund},\ and\ \citenamefont {Bulanov}}]{QED-review3}%
  \BibitemOpen
  \bibfield  {author} {\bibinfo {author} {\bibfnamefont {A.}~\bibnamefont {Gonoskov}}, \bibinfo {author} {\bibfnamefont {T.~G.}\ \bibnamefont {Blackburn}}, \bibinfo {author} {\bibfnamefont {M.}~\bibnamefont {Marklund}},\ and\ \bibinfo {author} {\bibfnamefont {S.~S.}\ \bibnamefont {Bulanov}},\ }\bibfield  {title} {\bibinfo {title} {Charged particle motion and radiation in strong electromagnetic fields},\ }\href {https://doi.org/10.1103/RevModPhys.94.045001} {\bibfield  {journal} {\bibinfo  {journal} {Rev. Mod. Phys.}\ }\textbf {\bibinfo {volume} {94}},\ \bibinfo {pages} {045001} (\bibinfo {year} {2022})}\BibitemShut {NoStop}%
\bibitem [{\citenamefont {Neitz}\ and\ \citenamefont {Di~Piazza}(2013)}]{Di-Piazza}%
  \BibitemOpen
  \bibfield  {author} {\bibinfo {author} {\bibfnamefont {N.}~\bibnamefont {Neitz}}\ and\ \bibinfo {author} {\bibfnamefont {A.}~\bibnamefont {Di~Piazza}},\ }\bibfield  {title} {\bibinfo {title} {Stochasticity effects in quantum radiation reaction},\ }\href {https://doi.org/10.1103/PhysRevLett.111.054802} {\bibfield  {journal} {\bibinfo  {journal} {Phys. Rev. Lett.}\ }\textbf {\bibinfo {volume} {111}},\ \bibinfo {pages} {054802} (\bibinfo {year} {2013})}\BibitemShut {NoStop}%
\bibitem [{\citenamefont {Griffith}\ \emph {et~al.}(2024)\citenamefont {Griffith}, \citenamefont {Qu},\ and\ \citenamefont {Fisch}}]{Fisch2024}%
  \BibitemOpen
  \bibfield  {author} {\bibinfo {author} {\bibfnamefont {A.}~\bibnamefont {Griffith}}, \bibinfo {author} {\bibfnamefont {K.}~\bibnamefont {Qu}},\ and\ \bibinfo {author} {\bibfnamefont {N.~J.}\ \bibnamefont {Fisch}},\ }\bibfield  {title} {\bibinfo {title} {Radiation reaction kinetics and collective qed signatures},\ }\href@noop {} {\bibfield  {journal} {\bibinfo  {journal} {Physics of Plasmas}\ }\textbf {\bibinfo {volume} {31}} (\bibinfo {year} {2024})}\BibitemShut {NoStop}%
\bibitem [{\citenamefont {Li}\ \emph {et~al.}(2023)\citenamefont {Li}, \citenamefont {Leng},\ and\ \citenamefont {Li}}]{li2023further}%
  \BibitemOpen
  \bibfield  {author} {\bibinfo {author} {\bibfnamefont {Z.}~\bibnamefont {Li}}, \bibinfo {author} {\bibfnamefont {Y.}~\bibnamefont {Leng}},\ and\ \bibinfo {author} {\bibfnamefont {R.}~\bibnamefont {Li}},\ }\bibfield  {title} {\bibinfo {title} {Further development of the short-pulse petawatt laser: trends, technologies, and bottlenecks},\ }\href@noop {} {\bibfield  {journal} {\bibinfo  {journal} {Laser \& Photonics Reviews}\ }\textbf {\bibinfo {volume} {17}},\ \bibinfo {pages} {2100705} (\bibinfo {year} {2023})}\BibitemShut {NoStop}%
\bibitem [{\citenamefont {Burke}\ \emph {et~al.}(1997)\citenamefont {Burke}, \citenamefont {Field}, \citenamefont {Horton-Smith}, \citenamefont {Spencer}, \citenamefont {Walz}, \citenamefont {Berridge}, \citenamefont {Bugg}, \citenamefont {Shmakov}, \citenamefont {Weidemann}, \citenamefont {Bula}, \citenamefont {McDonald}, \citenamefont {Prebys}, \citenamefont {Bamber}, \citenamefont {Boege}, \citenamefont {Koffas}, \citenamefont {Kotseroglou}, \citenamefont {Melissinos}, \citenamefont {Meyerhofer}, \citenamefont {Reis},\ and\ \citenamefont {Ragg}}]{SLAC1997}%
  \BibitemOpen
  \bibfield  {author} {\bibinfo {author} {\bibfnamefont {D.~L.}\ \bibnamefont {Burke}}, \bibinfo {author} {\bibfnamefont {R.~C.}\ \bibnamefont {Field}}, \bibinfo {author} {\bibfnamefont {G.}~\bibnamefont {Horton-Smith}}, \bibinfo {author} {\bibfnamefont {J.~E.}\ \bibnamefont {Spencer}}, \bibinfo {author} {\bibfnamefont {D.}~\bibnamefont {Walz}}, \bibinfo {author} {\bibfnamefont {S.~C.}\ \bibnamefont {Berridge}}, \bibinfo {author} {\bibfnamefont {W.~M.}\ \bibnamefont {Bugg}}, \bibinfo {author} {\bibfnamefont {K.}~\bibnamefont {Shmakov}}, \bibinfo {author} {\bibfnamefont {A.~W.}\ \bibnamefont {Weidemann}}, \bibinfo {author} {\bibfnamefont {C.}~\bibnamefont {Bula}}, \bibinfo {author} {\bibfnamefont {K.~T.}\ \bibnamefont {McDonald}}, \bibinfo {author} {\bibfnamefont {E.~J.}\ \bibnamefont {Prebys}}, \bibinfo {author} {\bibfnamefont {C.}~\bibnamefont {Bamber}}, \bibinfo {author} {\bibfnamefont {S.~J.}\ \bibnamefont {Boege}}, \bibinfo {author} {\bibfnamefont {T.}~\bibnamefont {Koffas}}, \bibinfo {author}
  {\bibfnamefont {T.}~\bibnamefont {Kotseroglou}}, \bibinfo {author} {\bibfnamefont {A.~C.}\ \bibnamefont {Melissinos}}, \bibinfo {author} {\bibfnamefont {D.~D.}\ \bibnamefont {Meyerhofer}}, \bibinfo {author} {\bibfnamefont {D.~A.}\ \bibnamefont {Reis}},\ and\ \bibinfo {author} {\bibfnamefont {W.}~\bibnamefont {Ragg}},\ }\bibfield  {title} {\bibinfo {title} {Positron production in multiphoton light-by-light scattering},\ }\href {https://doi.org/10.1103/PhysRevLett.79.1626} {\bibfield  {journal} {\bibinfo  {journal} {Phys. Rev. Lett.}\ }\textbf {\bibinfo {volume} {79}},\ \bibinfo {pages} {1626} (\bibinfo {year} {1997})}\BibitemShut {NoStop}%
\bibitem [{\citenamefont {Cole}\ \emph {et~al.}(2018)\citenamefont {Cole}, \citenamefont {Behm}, \citenamefont {Gerstmayr}, \citenamefont {Blackburn}, \citenamefont {Wood}, \citenamefont {Baird}, \citenamefont {Duff}, \citenamefont {Harvey}, \citenamefont {Ilderton}, \citenamefont {Joglekar}, \citenamefont {Krushelnick}, \citenamefont {Kuschel}, \citenamefont {Marklund}, \citenamefont {McKenna}, \citenamefont {Murphy}, \citenamefont {Poder}, \citenamefont {Ridgers}, \citenamefont {Samarin}, \citenamefont {Sarri}, \citenamefont {Symes}, \citenamefont {Thomas}, \citenamefont {Warwick}, \citenamefont {Zepf}, \citenamefont {Najmudin},\ and\ \citenamefont {Mangles}}]{Rutherford}%
  \BibitemOpen
  \bibfield  {author} {\bibinfo {author} {\bibfnamefont {J.~M.}\ \bibnamefont {Cole}}, \bibinfo {author} {\bibfnamefont {K.~T.}\ \bibnamefont {Behm}}, \bibinfo {author} {\bibfnamefont {E.}~\bibnamefont {Gerstmayr}}, \bibinfo {author} {\bibfnamefont {T.~G.}\ \bibnamefont {Blackburn}}, \bibinfo {author} {\bibfnamefont {J.~C.}\ \bibnamefont {Wood}}, \bibinfo {author} {\bibfnamefont {C.~D.}\ \bibnamefont {Baird}}, \bibinfo {author} {\bibfnamefont {M.~J.}\ \bibnamefont {Duff}}, \bibinfo {author} {\bibfnamefont {C.}~\bibnamefont {Harvey}}, \bibinfo {author} {\bibfnamefont {A.}~\bibnamefont {Ilderton}}, \bibinfo {author} {\bibfnamefont {A.~S.}\ \bibnamefont {Joglekar}}, \bibinfo {author} {\bibfnamefont {K.}~\bibnamefont {Krushelnick}}, \bibinfo {author} {\bibfnamefont {S.}~\bibnamefont {Kuschel}}, \bibinfo {author} {\bibfnamefont {M.}~\bibnamefont {Marklund}}, \bibinfo {author} {\bibfnamefont {P.}~\bibnamefont {McKenna}}, \bibinfo {author} {\bibfnamefont {C.~D.}\ \bibnamefont {Murphy}}, \bibinfo {author}
  {\bibfnamefont {K.}~\bibnamefont {Poder}}, \bibinfo {author} {\bibfnamefont {C.~P.}\ \bibnamefont {Ridgers}}, \bibinfo {author} {\bibfnamefont {G.~M.}\ \bibnamefont {Samarin}}, \bibinfo {author} {\bibfnamefont {G.}~\bibnamefont {Sarri}}, \bibinfo {author} {\bibfnamefont {D.~R.}\ \bibnamefont {Symes}}, \bibinfo {author} {\bibfnamefont {A.~G.~R.}\ \bibnamefont {Thomas}}, \bibinfo {author} {\bibfnamefont {J.}~\bibnamefont {Warwick}}, \bibinfo {author} {\bibfnamefont {M.}~\bibnamefont {Zepf}}, \bibinfo {author} {\bibfnamefont {Z.}~\bibnamefont {Najmudin}},\ and\ \bibinfo {author} {\bibfnamefont {S.~P.~D.}\ \bibnamefont {Mangles}},\ }\bibfield  {title} {\bibinfo {title} {Experimental evidence of radiation reaction in the collision of a high-intensity laser pulse with a laser-wakefield accelerated electron beam},\ }\href {https://doi.org/10.1103/PhysRevX.8.011020} {\bibfield  {journal} {\bibinfo  {journal} {Phys. Rev. X}\ }\textbf {\bibinfo {volume} {8}},\ \bibinfo {pages} {011020} (\bibinfo {year}
  {2018})}\BibitemShut {NoStop}%
\bibitem [{\citenamefont {Chen}\ and\ \citenamefont {Fiuza}(2023)}]{Fiuza2023}%
  \BibitemOpen
  \bibfield  {author} {\bibinfo {author} {\bibfnamefont {H.}~\bibnamefont {Chen}}\ and\ \bibinfo {author} {\bibfnamefont {F.}~\bibnamefont {Fiuza}},\ }\bibfield  {title} {\bibinfo {title} {Perspectives on relativistic electron--positron pair plasma experiments of astrophysical relevance using high-power lasers},\ }\href@noop {} {\bibfield  {journal} {\bibinfo  {journal} {Physics of Plasmas}\ }\textbf {\bibinfo {volume} {30}} (\bibinfo {year} {2023})}\BibitemShut {NoStop}%
\bibitem [{\citenamefont {Landau}(2013)}]{Landau_Lif}%
  \BibitemOpen
  \bibfield  {author} {\bibinfo {author} {\bibfnamefont {L.~D.}\ \bibnamefont {Landau}},\ }\href@noop {} {\emph {\bibinfo {title} {The classical theory of fields}}},\ Vol.~\bibinfo {volume} {2}\ (\bibinfo  {publisher} {Elsevier},\ \bibinfo {year} {2013})\BibitemShut {NoStop}%
\bibitem [{\citenamefont {Jackson}(1999)}]{Jackson}%
  \BibitemOpen
  \bibfield  {author} {\bibinfo {author} {\bibfnamefont {J.~D.}\ \bibnamefont {Jackson}},\ }\bibfield  {title} {\bibinfo {title} {Classical electrodynamics john wiley \& sons},\ }\href@noop {} {\bibfield  {journal} {\bibinfo  {journal} {Inc., New York}\ }\textbf {\bibinfo {volume} {13}} (\bibinfo {year} {1999})}\BibitemShut {NoStop}%
\bibitem [{\citenamefont {Burton}\ and\ \citenamefont {Noble}(2014)}]{Burton}%
  \BibitemOpen
  \bibfield  {author} {\bibinfo {author} {\bibfnamefont {D.~A.}\ \bibnamefont {Burton}}\ and\ \bibinfo {author} {\bibfnamefont {A.}~\bibnamefont {Noble}},\ }\bibfield  {title} {\bibinfo {title} {Aspects of electromagnetic radiation reaction in strong fields},\ }\href@noop {} {\bibfield  {journal} {\bibinfo  {journal} {Contemporary Physics}\ }\textbf {\bibinfo {volume} {55}},\ \bibinfo {pages} {110} (\bibinfo {year} {2014})}\BibitemShut {NoStop}%
\bibitem [{\citenamefont {Bulanov}\ \emph {et~al.}(2011)\citenamefont {Bulanov}, \citenamefont {Esirkepov}, \citenamefont {Kando}, \citenamefont {Koga},\ and\ \citenamefont {Bulanov}}]{Bulanov}%
  \BibitemOpen
  \bibfield  {author} {\bibinfo {author} {\bibfnamefont {S.~V.}\ \bibnamefont {Bulanov}}, \bibinfo {author} {\bibfnamefont {T.~Z.}\ \bibnamefont {Esirkepov}}, \bibinfo {author} {\bibfnamefont {M.}~\bibnamefont {Kando}}, \bibinfo {author} {\bibfnamefont {J.~K.}\ \bibnamefont {Koga}},\ and\ \bibinfo {author} {\bibfnamefont {S.~S.}\ \bibnamefont {Bulanov}},\ }\bibfield  {title} {\bibinfo {title} {Lorentz-abraham-dirac versus landau-lifshitz radiation friction force in the ultrarelativistic electron interaction with electromagnetic wave (exact solutions)},\ }\href {https://doi.org/10.1103/PhysRevE.84.056605} {\bibfield  {journal} {\bibinfo  {journal} {Phys. Rev. E}\ }\textbf {\bibinfo {volume} {84}},\ \bibinfo {pages} {056605} (\bibinfo {year} {2011})}\BibitemShut {NoStop}%
\bibitem [{\citenamefont {Vranic}\ \emph {et~al.}(2016{\natexlab{a}})\citenamefont {Vranic}, \citenamefont {Martins}, \citenamefont {Fonseca},\ and\ \citenamefont {Silva}}]{Silva}%
  \BibitemOpen
  \bibfield  {author} {\bibinfo {author} {\bibfnamefont {M.}~\bibnamefont {Vranic}}, \bibinfo {author} {\bibfnamefont {J.~L.}\ \bibnamefont {Martins}}, \bibinfo {author} {\bibfnamefont {R.~A.}\ \bibnamefont {Fonseca}},\ and\ \bibinfo {author} {\bibfnamefont {L.~O.}\ \bibnamefont {Silva}},\ }\bibfield  {title} {\bibinfo {title} {Classical radiation reaction in particle-in-cell simulations},\ }\href@noop {} {\bibfield  {journal} {\bibinfo  {journal} {Computer Physics Communications}\ }\textbf {\bibinfo {volume} {204}},\ \bibinfo {pages} {141} (\bibinfo {year} {2016}{\natexlab{a}})}\BibitemShut {NoStop}%
\bibitem [{\citenamefont {Wallin}\ \emph {et~al.}(2017)\citenamefont {Wallin}, \citenamefont {Gonoskov}, \citenamefont {Harvey}, \citenamefont {Lundh},\ and\ \citenamefont {Marklund}}]{Wallin}%
  \BibitemOpen
  \bibfield  {author} {\bibinfo {author} {\bibfnamefont {E.}~\bibnamefont {Wallin}}, \bibinfo {author} {\bibfnamefont {A.}~\bibnamefont {Gonoskov}}, \bibinfo {author} {\bibfnamefont {C.}~\bibnamefont {Harvey}}, \bibinfo {author} {\bibfnamefont {O.}~\bibnamefont {Lundh}},\ and\ \bibinfo {author} {\bibfnamefont {M.}~\bibnamefont {Marklund}},\ }\bibfield  {title} {\bibinfo {title} {Ultra-intense laser pulses in near-critical underdense plasmas--radiation reaction and energy partitioning},\ }\href@noop {} {\bibfield  {journal} {\bibinfo  {journal} {Journal of Plasma Physics}\ }\textbf {\bibinfo {volume} {83}} (\bibinfo {year} {2017})}\BibitemShut {NoStop}%
\bibitem [{\citenamefont {Vranic}\ \emph {et~al.}(2016{\natexlab{b}})\citenamefont {Vranic}, \citenamefont {Grismayer}, \citenamefont {Fonseca},\ and\ \citenamefont {Silva}}]{Silva_Q}%
  \BibitemOpen
  \bibfield  {author} {\bibinfo {author} {\bibfnamefont {M.}~\bibnamefont {Vranic}}, \bibinfo {author} {\bibfnamefont {T.}~\bibnamefont {Grismayer}}, \bibinfo {author} {\bibfnamefont {R.~A.}\ \bibnamefont {Fonseca}},\ and\ \bibinfo {author} {\bibfnamefont {L.~O.}\ \bibnamefont {Silva}},\ }\bibfield  {title} {\bibinfo {title} {Quantum radiation reaction in head-on laser-electron beam interaction},\ }\href@noop {} {\bibfield  {journal} {\bibinfo  {journal} {New Journal of Physics}\ }\textbf {\bibinfo {volume} {18}},\ \bibinfo {pages} {073035} (\bibinfo {year} {2016}{\natexlab{b}})}\BibitemShut {NoStop}%
\bibitem [{\citenamefont {Chen}\ \emph {et~al.}(2022)\citenamefont {Chen}, \citenamefont {Meuren}, \citenamefont {Gerstmayr}, \citenamefont {Yakimenko}, \citenamefont {Bucksbaum},\ and\ \citenamefont {Reis}}]{E320}%
  \BibitemOpen
  \bibfield  {author} {\bibinfo {author} {\bibfnamefont {Z.}~\bibnamefont {Chen}}, \bibinfo {author} {\bibfnamefont {S.}~\bibnamefont {Meuren}}, \bibinfo {author} {\bibfnamefont {E.}~\bibnamefont {Gerstmayr}}, \bibinfo {author} {\bibfnamefont {V.}~\bibnamefont {Yakimenko}}, \bibinfo {author} {\bibfnamefont {P.~H.}\ \bibnamefont {Bucksbaum}},\ and\ \bibinfo {author} {\bibfnamefont {D.~A.}\ \bibnamefont {Reis}},\ }\bibfield  {title} {\bibinfo {title} {Preparation of strong-field qed experiments at facet-ii},\ }in\ \href@noop {} {\emph {\bibinfo {booktitle} {High Intensity Lasers and High Field Phenomena}}}\ (\bibinfo {organization} {Optica Publishing Group},\ \bibinfo {year} {2022})\ pp.\ \bibinfo {pages} {HF4B--6}\BibitemShut {NoStop}%
\bibitem [{\citenamefont {Abramowicz}\ \emph {et~al.}(2021{\natexlab{a}})\citenamefont {Abramowicz}, \citenamefont {Acosta}, \citenamefont {Altarelli}, \citenamefont {Assmann}, \citenamefont {Bai}, \citenamefont {Behnke}, \citenamefont {Benhammou}, \citenamefont {Blackburn}, \citenamefont {Boogert}, \citenamefont {Borysov} \emph {et~al.}}]{LUXE}%
  \BibitemOpen
  \bibfield  {author} {\bibinfo {author} {\bibfnamefont {H.}~\bibnamefont {Abramowicz}}, \bibinfo {author} {\bibfnamefont {U.}~\bibnamefont {Acosta}}, \bibinfo {author} {\bibfnamefont {M.}~\bibnamefont {Altarelli}}, \bibinfo {author} {\bibfnamefont {R.}~\bibnamefont {Assmann}}, \bibinfo {author} {\bibfnamefont {Z.}~\bibnamefont {Bai}}, \bibinfo {author} {\bibfnamefont {T.}~\bibnamefont {Behnke}}, \bibinfo {author} {\bibfnamefont {Y.}~\bibnamefont {Benhammou}}, \bibinfo {author} {\bibfnamefont {T.}~\bibnamefont {Blackburn}}, \bibinfo {author} {\bibfnamefont {S.}~\bibnamefont {Boogert}}, \bibinfo {author} {\bibfnamefont {O.}~\bibnamefont {Borysov}}, \emph {et~al.},\ }\bibfield  {title} {\bibinfo {title} {Conceptual design report for the luxe experiment},\ }\href@noop {} {\bibfield  {journal} {\bibinfo  {journal} {The European Physical Journal Special Topics}\ ,\ \bibinfo {pages} {1}} (\bibinfo {year} {2021}{\natexlab{a}})}\BibitemShut {NoStop}%
\bibitem [{\citenamefont {Ritus}(1985)}]{Ritus1985}%
  \BibitemOpen
  \bibfield  {author} {\bibinfo {author} {\bibfnamefont {V.}~\bibnamefont {Ritus}},\ }\bibfield  {title} {\bibinfo {title} {Quantum effects of the interaction of elementary particles with an intense electromagnetic field},\ }\href@noop {} {\bibfield  {journal} {\bibinfo  {journal} {J. Sov. Laser Res.;(United States)}\ }\textbf {\bibinfo {volume} {6}} (\bibinfo {year} {1985})}\BibitemShut {NoStop}%
\bibitem [{\citenamefont {Abramowicz}\ \emph {et~al.}(2021{\natexlab{b}})\citenamefont {Abramowicz}, \citenamefont {Acosta}, \citenamefont {Altarelli}, \citenamefont {Assmann}, \citenamefont {Bai}, \citenamefont {Behnke}, \citenamefont {Benhammou}, \citenamefont {Blackburn}, \citenamefont {Boogert}, \citenamefont {Borysov} \emph {et~al.}}]{QED-review2}%
  \BibitemOpen
  \bibfield  {author} {\bibinfo {author} {\bibfnamefont {H.}~\bibnamefont {Abramowicz}}, \bibinfo {author} {\bibfnamefont {U.}~\bibnamefont {Acosta}}, \bibinfo {author} {\bibfnamefont {M.}~\bibnamefont {Altarelli}}, \bibinfo {author} {\bibfnamefont {R.}~\bibnamefont {Assmann}}, \bibinfo {author} {\bibfnamefont {Z.}~\bibnamefont {Bai}}, \bibinfo {author} {\bibfnamefont {T.}~\bibnamefont {Behnke}}, \bibinfo {author} {\bibfnamefont {Y.}~\bibnamefont {Benhammou}}, \bibinfo {author} {\bibfnamefont {T.}~\bibnamefont {Blackburn}}, \bibinfo {author} {\bibfnamefont {S.}~\bibnamefont {Boogert}}, \bibinfo {author} {\bibfnamefont {O.}~\bibnamefont {Borysov}}, \emph {et~al.},\ }\bibfield  {title} {\bibinfo {title} {Conceptual design report for the luxe experiment},\ }\href@noop {} {\bibfield  {journal} {\bibinfo  {journal} {The European Physical Journal Special Topics}\ }\textbf {\bibinfo {volume} {230}},\ \bibinfo {pages} {2445} (\bibinfo {year} {2021}{\natexlab{b}})}\BibitemShut {NoStop}%
\bibitem [{\citenamefont {Brodin}\ \emph {et~al.}(2022)\citenamefont {Brodin}, \citenamefont {Al-Naseri}, \citenamefont {Zamanian}, \citenamefont {Torgrimsson},\ and\ \citenamefont {Eliasson}}]{E-schwinger}%
  \BibitemOpen
  \bibfield  {author} {\bibinfo {author} {\bibfnamefont {G.}~\bibnamefont {Brodin}}, \bibinfo {author} {\bibfnamefont {H.}~\bibnamefont {Al-Naseri}}, \bibinfo {author} {\bibfnamefont {J.}~\bibnamefont {Zamanian}}, \bibinfo {author} {\bibfnamefont {G.}~\bibnamefont {Torgrimsson}},\ and\ \bibinfo {author} {\bibfnamefont {B.}~\bibnamefont {Eliasson}},\ }\bibfield  {title} {\bibinfo {title} {Plasma dynamics at the schwinger limit and beyond},\ }\href@noop {} {\bibfield  {journal} {\bibinfo  {journal} {arXiv preprint arXiv:2209.07872}\ } (\bibinfo {year} {2022})}\BibitemShut {NoStop}%
\bibitem [{\citenamefont {Hakim}\ and\ \citenamefont {Mangeney}(1968)}]{Hakim}%
  \BibitemOpen
  \bibfield  {author} {\bibinfo {author} {\bibfnamefont {R.}~\bibnamefont {Hakim}}\ and\ \bibinfo {author} {\bibfnamefont {A.}~\bibnamefont {Mangeney}},\ }\bibfield  {title} {\bibinfo {title} {Relativistic kinetic equations including radiation effects. i. vlasov approximation},\ }\href@noop {} {\bibfield  {journal} {\bibinfo  {journal} {Journal of Mathematical Physics}\ }\textbf {\bibinfo {volume} {9}},\ \bibinfo {pages} {116} (\bibinfo {year} {1968})}\BibitemShut {NoStop}%
\bibitem [{\citenamefont {Kunze}\ and\ \citenamefont {Rendall}(2001)}]{Kunze}%
  \BibitemOpen
  \bibfield  {author} {\bibinfo {author} {\bibfnamefont {M.}~\bibnamefont {Kunze}}\ and\ \bibinfo {author} {\bibfnamefont {A.~D.}\ \bibnamefont {Rendall}},\ }\bibfield  {title} {\bibinfo {title} {The vlasov-poisson system with radiation damping},\ }in\ \href@noop {} {\emph {\bibinfo {booktitle} {Annales Henri Poincar{\'e}}}},\ Vol.~\bibinfo {volume} {2}\ (\bibinfo {organization} {Springer},\ \bibinfo {year} {2001})\ pp.\ \bibinfo {pages} {857--886}\BibitemShut {NoStop}%
\bibitem [{\citenamefont {Elskens}\ and\ \citenamefont {Kiessling}(2020)}]{Elskens}%
  \BibitemOpen
  \bibfield  {author} {\bibinfo {author} {\bibfnamefont {Y.}~\bibnamefont {Elskens}}\ and\ \bibinfo {author} {\bibfnamefont {M.-H.}\ \bibnamefont {Kiessling}},\ }\bibfield  {title} {\bibinfo {title} {Microscopic foundations of kinetic plasma theory: The relativistic vlasov--maxwell equations and their radiation-reaction-corrected generalization},\ }\href@noop {} {\bibfield  {journal} {\bibinfo  {journal} {Journal of Statistical Physics}\ }\textbf {\bibinfo {volume} {180}},\ \bibinfo {pages} {749} (\bibinfo {year} {2020})}\BibitemShut {NoStop}%
\bibitem [{\citenamefont {Noble}\ \emph {et~al.}()\citenamefont {Noble}, \citenamefont {Gratus}, \citenamefont {Burton}, \citenamefont {Ersfeld}, \citenamefont {Islam} \emph {et~al.}}]{Burton-2}%
  \BibitemOpen
  \bibfield  {author} {\bibinfo {author} {\bibfnamefont {A.}~\bibnamefont {Noble}}, \bibinfo {author} {\bibfnamefont {J.}~\bibnamefont {Gratus}}, \bibinfo {author} {\bibfnamefont {D.}~\bibnamefont {Burton}}, \bibinfo {author} {\bibfnamefont {B.}~\bibnamefont {Ersfeld}}, \bibinfo {author} {\bibfnamefont {M.~R.}\ \bibnamefont {Islam}}, \emph {et~al.},\ }\bibfield  {title} {\bibinfo {title} {Kinetic treatment of radiation reaction effects},\ }in\ \href@noop {} {\emph {\bibinfo {booktitle} {Proc. of SPIE Vol}}},\ Vol.\ \bibinfo {volume} {8079},\ pp.\ \bibinfo {pages} {80790L--1}\BibitemShut {NoStop}%
\bibitem [{\citenamefont {Berezhiani}\ \emph {et~al.}(2004)\citenamefont {Berezhiani}, \citenamefont {Hazeltine},\ and\ \citenamefont {Mahajan}}]{Mahajan1}%
  \BibitemOpen
  \bibfield  {author} {\bibinfo {author} {\bibfnamefont {V.~I.}\ \bibnamefont {Berezhiani}}, \bibinfo {author} {\bibfnamefont {R.~D.}\ \bibnamefont {Hazeltine}},\ and\ \bibinfo {author} {\bibfnamefont {S.~M.}\ \bibnamefont {Mahajan}},\ }\bibfield  {title} {\bibinfo {title} {Radiation reaction and relativistic hydrodynamics},\ }\href {https://doi.org/10.1103/PhysRevE.69.056406} {\bibfield  {journal} {\bibinfo  {journal} {Phys. Rev. E}\ }\textbf {\bibinfo {volume} {69}},\ \bibinfo {pages} {056406} (\bibinfo {year} {2004})}\BibitemShut {NoStop}%
\bibitem [{\citenamefont {Hazeltine}\ and\ \citenamefont {Mahajan}(2004)}]{Mahajan2}%
  \BibitemOpen
  \bibfield  {author} {\bibinfo {author} {\bibfnamefont {R.~D.}\ \bibnamefont {Hazeltine}}\ and\ \bibinfo {author} {\bibfnamefont {S.~M.}\ \bibnamefont {Mahajan}},\ }\bibfield  {title} {\bibinfo {title} {Closed fluid description of relativistic, magnetized plasma interacting with radiation field},\ }\href {https://doi.org/10.1103/PhysRevE.70.036404} {\bibfield  {journal} {\bibinfo  {journal} {Phys. Rev. E}\ }\textbf {\bibinfo {volume} {70}},\ \bibinfo {pages} {036404} (\bibinfo {year} {2004})}\BibitemShut {NoStop}%
\bibitem [{\citenamefont {Berezhiani}\ \emph {et~al.}(2008)\citenamefont {Berezhiani}, \citenamefont {Mahajan},\ and\ \citenamefont {Yoshida}}]{Mahajan3}%
  \BibitemOpen
  \bibfield  {author} {\bibinfo {author} {\bibfnamefont {V.~I.}\ \bibnamefont {Berezhiani}}, \bibinfo {author} {\bibfnamefont {S.~M.}\ \bibnamefont {Mahajan}},\ and\ \bibinfo {author} {\bibfnamefont {Z.}~\bibnamefont {Yoshida}},\ }\bibfield  {title} {\bibinfo {title} {Plasma acceleration and cooling by strong laser field due to the action of radiation reaction force},\ }\href {https://doi.org/10.1103/PhysRevE.78.066403} {\bibfield  {journal} {\bibinfo  {journal} {Phys. Rev. E}\ }\textbf {\bibinfo {volume} {78}},\ \bibinfo {pages} {066403} (\bibinfo {year} {2008})}\BibitemShut {NoStop}%
\bibitem [{\citenamefont {Dalakishvili}\ \emph {et~al.}(2007)\citenamefont {Dalakishvili}, \citenamefont {Rogava},\ and\ \citenamefont {Berezhiani}}]{Dalakishvili}%
  \BibitemOpen
  \bibfield  {author} {\bibinfo {author} {\bibfnamefont {G.~T.}\ \bibnamefont {Dalakishvili}}, \bibinfo {author} {\bibfnamefont {A.~D.}\ \bibnamefont {Rogava}},\ and\ \bibinfo {author} {\bibfnamefont {V.~I.}\ \bibnamefont {Berezhiani}},\ }\bibfield  {title} {\bibinfo {title} {Role of radiation reaction forces in the dynamics of centrifugally accelerated particles},\ }\href {https://doi.org/10.1103/PhysRevD.76.045003} {\bibfield  {journal} {\bibinfo  {journal} {Phys. Rev. D}\ }\textbf {\bibinfo {volume} {76}},\ \bibinfo {pages} {045003} (\bibinfo {year} {2007})}\BibitemShut {NoStop}%
\bibitem [{\citenamefont {Al-Naseri}\ and\ \citenamefont {Brodin}(2023)}]{RR2023}%
  \BibitemOpen
  \bibfield  {author} {\bibinfo {author} {\bibfnamefont {H.}~\bibnamefont {Al-Naseri}}\ and\ \bibinfo {author} {\bibfnamefont {G.}~\bibnamefont {Brodin}},\ }\bibfield  {title} {\bibinfo {title} {Radiation reaction effects in relativistic plasmas: The electrostatic limit},\ }\href@noop {} {\bibfield  {journal} {\bibinfo  {journal} {Physical Review E}\ }\textbf {\bibinfo {volume} {107}},\ \bibinfo {pages} {035203} (\bibinfo {year} {2023})}\BibitemShut {NoStop}%
\bibitem [{\citenamefont {Baier}\ \emph {et~al.}(1998)\citenamefont {Baier}, \citenamefont {Katkov},\ and\ \citenamefont {Strakhovenko}}]{QEDProbabilities}%
  \BibitemOpen
  \bibfield  {author} {\bibinfo {author} {\bibfnamefont {V.~N.}\ \bibnamefont {Baier}}, \bibinfo {author} {\bibfnamefont {V.~M.}\ \bibnamefont {Katkov}},\ and\ \bibinfo {author} {\bibfnamefont {V.~M.}\ \bibnamefont {Strakhovenko}},\ }\href@noop {} {\emph {\bibinfo {title} {Electromagnetic processes at high energies in oriented single crystals}}}\ (\bibinfo  {publisher} {World Scientific},\ \bibinfo {year} {1998})\BibitemShut {NoStop}%
\end{thebibliography}%

\end{document}